International
Virtual
Observatory
Alliance

# IVOA Spectrum Data Model

## Version 1.1

## IVOA Recommendation 20 November 2011

**Editors:**
Jonathan McDowell, Doug Tody
**Contributors:**
Jonathan McDowell, Doug Tody, Tamas Budavari, Markus Dolensky, Inga Kamp, Kelly McCusker, Pavlos Protopapas, Arnold Rots, Randy Thompson, Frank Valdes, Petr Skoda, Bruno Rino, Sebastien Derriére, Jesus Salgado, Omar Laurino and the IVOA Data Access Layer and Data Model Working Groups.

**Abstract**

We present a data model describing the structure of spectrophotometric datasets with spectral and temporal coordinates and associated metadata. This data model may be used to represent spectra, time series data, segments of SED (Spectral Energy Distributions) and other spectral or temporal associations.

## Status of this document

This document has been produced by the Data Model Working Group. It has been reviewed by IVOA Members and other interested parties, and has been endorsed by the IVOA Executive Committee as an IVOA Recommendation. It is a stable document and may be used as reference material or cited as a normative reference from another document. IVOA's role in making the Recommendation is to draw attention to the specification and to promote its widespread deployment. This enhances the functionality and interoperability inside the Astronomical Community.

This document has been developed with support from the National Science Foundation's[1] Information Technology Research Program under Cooperative Agreement AST0122449 with The Johns Hopkins University, from the UK Particle Physics and Astronomy Research Council (PPARC[2]) and from the European Commission's Sixth Framework Program[3] via the Optical Infrared Coordination Network (OPTICON[4]).

The **Virtual Observatory (VO)** is general term for a collection of federated resources that can be used to conduct astronomical research, education, and outreach.

The **International Virtual Observatory Alliance (IVOA)** (`http://www.ivoa.net`) is a global collaboration of separately funded projects to develop standards and infrastructure that enable VO applications.

[1] http://www.nsf.gov/
[2] http://www.pparc.ac.uk
[3] http://fp6.cordis.lu/fp6/home.cfm
[4] http://www.astro-opticon.org

# Contents

# Part 1: Spectrum Data Model

# 1 Introduction and Motivation

Spectra are stored in many different ways within the astronomical community. In this document we present a proposed abstraction for spectral data, which can be used for describing spectrum datasets and also reused by other standards. We also provide serializations for the spectrum dataset use case in VOTABLE, FITS, and XML, for use as a standard method of spectral data interchange, and define mandatory, recommended and optional fields for that use case.

We distinguish in several places between the implementation proposed in this document, referred to as Version 1, and capabilities proposed for possible later implementation.

This version fixes some issues coming from implementation feedback. Most fixes are minor and do not represent any fundamental changes with respect to the previous Recommendation. Modifications to UCDs (Table 1) can have an impact on implementations and/or validators.

## 1.1 Change Log

```
2011 Oct 20 V1.1 Mod 8 - Section 1 (Motivation): added scope of v1.1 changes.
  Section 1.2 (Architecture): dropped references to SpectrumDM1.2.
  References update. (As suggested by Christophe Arviset).

2011 Oct 17  V1.1 Mod 7 - UCD updates suggested by S. Derriere.

2011 Mar 19  V1.1 Mod 6  - Architecture section

2011 Mar 3   V1.1 (1.04) Mod 5 - Typo corrections

2011 Jan 19  V1.04 Mod 4 - Further minor updates for 2009 Dec B. Rino comments.

2010 Dec 19  V1.04 Mod 3 - Further text clarifications to the previous changes.
Note that there have been no changes to the XML schema.

2010 Apr 29  V1.04 Mod 2 Updated to account for comments by B. Rino.
 - added extra defatuls for missing utypes
  in FITS (CoordSys.SpaceFrame.UCD, CoordSys.TimeFrame.UCD, CoordSys.SpectralFrame.UCD).
 - added FITS SKY_REF keyword for missing utype CoordSys.SpaceFrame.RefPos
 - clarified meaning of '*' shorthand in listings of UCDs.

2010 Apr 21  V1.04: Made clear that 'mandatory' is for the spectrum document use case
and that other use cases can redfine which items are mandatory.

2009 Apr  5  V1.03 Mod 1: Fixed typo for ESAC SI dimensional codes and corrected stat.error
UCDs (stat.error must be primary).

2007 Oct 29 V1.03: SpectrumDM 1.03 released as a Recommendation

2007 Oct 28  V1.02 Rev 2: Corrected HTML version; corrected author list

2007 Aug 26  V1.01 RC3 Rev 2 = V1.02 RC3 Rev 2
 - Minor edits and corrections to examples;
```

```
clarification of IVOA identifiers; added NORMALIZED
calibration option. (Date changed to Sep 13 for IVOA Doc submission only;
version number changed at request of IVOA Doc Coord)

2007 Jul 10  V1.01 RC3 Rev 1
 - Minor edits following RFC

2007 May 15  V1.01 RC2 Rev 15
 - Proposed recommendation.

2007 May  1  V1.01 RC2 Rev 14
 - Trivial cover formatting

2007 Apr 30 V1.01 RC2 Rev 13
 - Added Support.Extent as suggested by A. Micol
 - Improved text in several places

2007 Apr 26 V1.01 RC2 Rev 12
 - UCD time.expo updated to time.duration/stop/start;obs.exposure
 - Updated VOTABLE examples

2007 Apr 25 V1.01 RC2 Rev 11
 - Included the correct file; Rev 10 was bogus

2007 Apr 17 V1.01 RC2 Rev 10
 - Fixed errors in XSD and in text
 - Revised Characterization text
 - Added REST_Z FITS keyword for CoordSys.SpectralFrame.Redshift

2007 Apr 12 V1.01 RC2 Rev 9
 - Incorporate D Tody comments
 - DataID.Title mandatory
 - Changes to recommended case of Utype fields e.g. Redshift not redshift.
 - Utypes involving stat.error changed to put the stat.error first.

2007 Apr 4 V1.01 RC2 Rev 8
 - FITS keyword TMID added; TDMINn/TDMAXn

2007 Apr 1 V1.01 RC2 Rev 7
 - Modifications for compatibility with Char working draft:
   Moved Calibration utype from CharAxis.Accuracy to CharAxis
   Added SamplingPrecision.SampleExtent and
   SamplingPrecision.SamplingPrecisionRefVal.FillFactor


2007 Feb 12 V1.01 RC2 Rev 5
 - Curation.Reference can have multiple instances
```

```
2007 Jan 17 V1.01 RC2 Rev 4
 - Changed FITS keyword SIZE to DATALEN (D Tody request)
 - Added text describing use of non standard units.
- Reformat units in Tables 2,3 to OGIP convention

2006 Dec 11 V1.01 RC2 Rev 3
 - Fixd more typos in XSD and XML example

2006 Dec 6 V1.01 RC2 Rev 2
 - Upgraded UCDs to version 1.21
 - Added SpectralAxis.ResPower and SPEC_RP keyword
   for resolving power; added element to XSD.
 - XSD changed segmentType definition to put Data element at end of sequence.
 - XSD corrected type errors in a few cases in Curation type.
 - XSD added missing elements CreatorDID, Bandpass to DataID.
 - XML instance example corrected errors in Characterization axes.
 - FITS keywords changed: CREATOR to AUTHOR; DER_ERR to DER_ZERR
 - FITS added more TUTYPn keyword examples.
 - FITS added comment on VOCSID
 - Corrected mistakes in FITS and VOT examples
 - Clarified role of Aperture
 - Further clarified CreatorDID, PublisherDID, DatasetID distinction.
 - Clarifications and corrections in text

2006 Oct 22 V1.0 RC1 (since V0.98d Rev 4)
- Added table numbers
- Changed some defaults in Table 1
- Added flux UCDs for transmission curves, polarized flux
- Amplified discussion of RedshiftFrame
- Added Spectral location and bounds
- Reorganized order of some sections
- Further rationalization of FITS keywords, rewrote FITS section
- Added TUCDn and TUTYPn
```

## 1.2 IVOA Architecture Context

Data Models in the VO aim to define the common elements of astronomical data and metadata collections and to provide a framework for describing their relationships so these become inter operable in a transparent manner. The Spectrum Data Model (SpectrumDM) standard presents a data model describing the structure of spectrophotometric datasets with spectral and temporal coordinates and associated metadata. This data model may be used to represent spectra, time series data, segments of SED (Spectral Energy Distributions) and other spectral or temporal associations. SpectrumDM is used with the associated Data Access Protocol, SSAP (Simple Spectra Access Protocol).

As with most of the VO Data Models, SpectrumDM makes use of STC, Utypes, Units and UCDs. Furthermore, SpectrumDM makes reference to the CharDM (Characterization Data

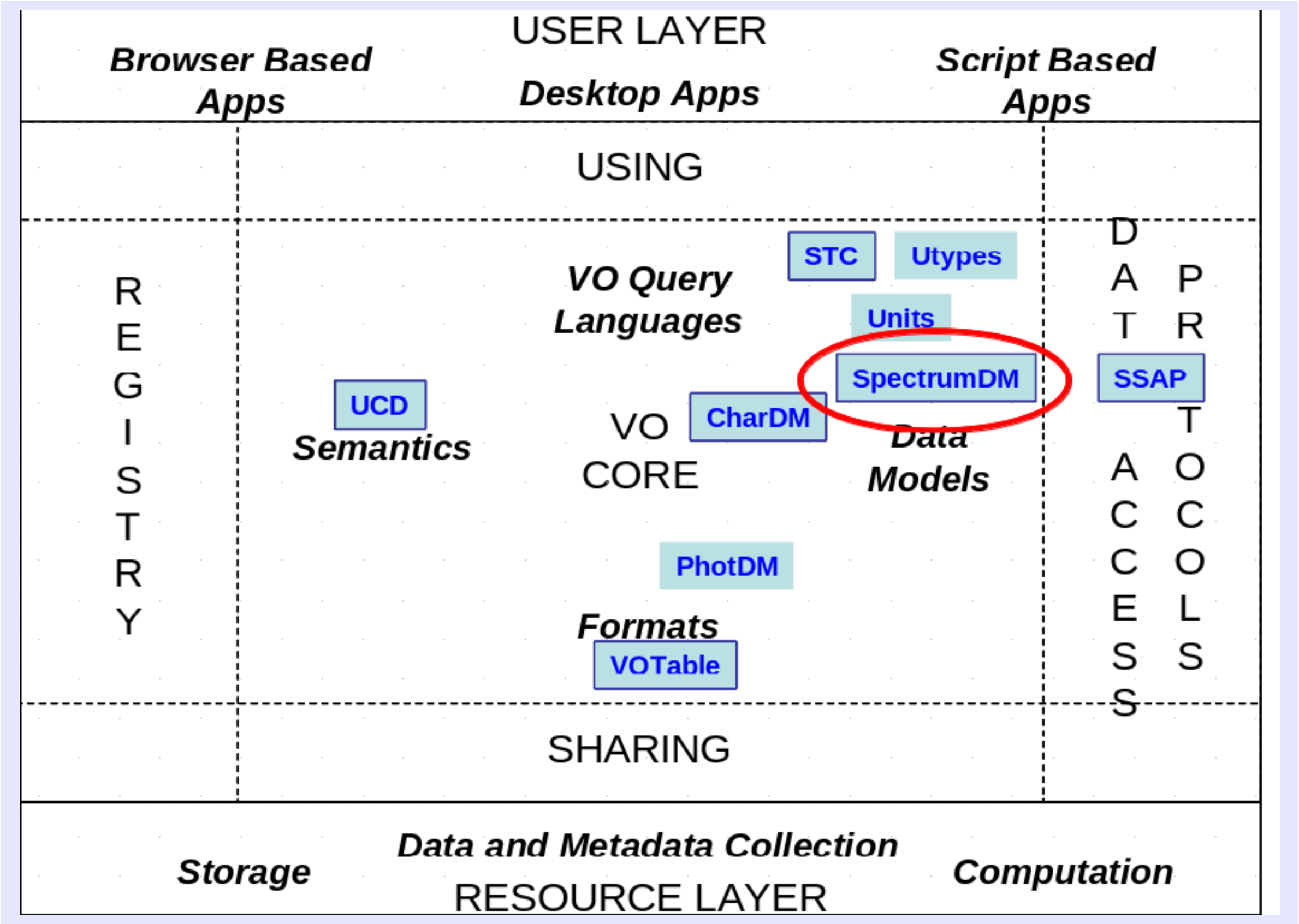

Figure 1: Spectrum DM in IVOA architecture

Model). It can be serialized with a VOTable, among other formats.

## 2 Requirements

We need to represent a single 1-dimensional spectrum in sufficient detail to understand the differences between two spectra of the same object and between two spectra of different objects.

We need to represent time series photometry, with many photometry points of the same object at different times.

Finally, we need to represent associations of spectra, such as the segments of an echelle spectrum, or spectral energy distributions (SED) which consist of multiple spectra and photometry points, usually for a single object. The 'SED' model will be described in a separate document which builds on the structures described here.

## 3 Spectral Data Model summary

### 3.1 Model Components

Our model for a spectrum is a set of one or more data points (photometry) each of which share the same contextual metadata (aperture, position, etc.). Specifically, a spectrum will have arrays of the following values:

- Flux value, with upper and lower statistical (uncorrelated) errors
- Spectral coordinate (e.g. wavelength), central and bin min and max
- (Optionally) Time coordinate, convertible to MJD UTC
- Optional Quality mask
- Optional spectral resolution array

and will have associated metadata including, for example,

- Data collection and Dataset ID
- Exposure time in seconds
- Position of aperture center, given as ICRS degrees (similar to J2000)
- Aperture size in degrees
- Systematic (correlated) error
- Bibcode

In later sections we elaborate these concepts in detail, including some complications that we explicitly do not attempt to handle in this version.

### 3.2 Use Cases and Required Fields

The main use case considered by this standard is the representation of a dataset (document) actually containing a spectrum. For this use case, the data model fields and possible values are listed. We distinguish between optional and required fields in the text, as well as via a column in the tables which has values of MAN (Mandatory, i.e. required), REC (Recommended) and OPT (Optional). Where appropriate we list those values of the physical units which interoperable implementations are required to recognize.

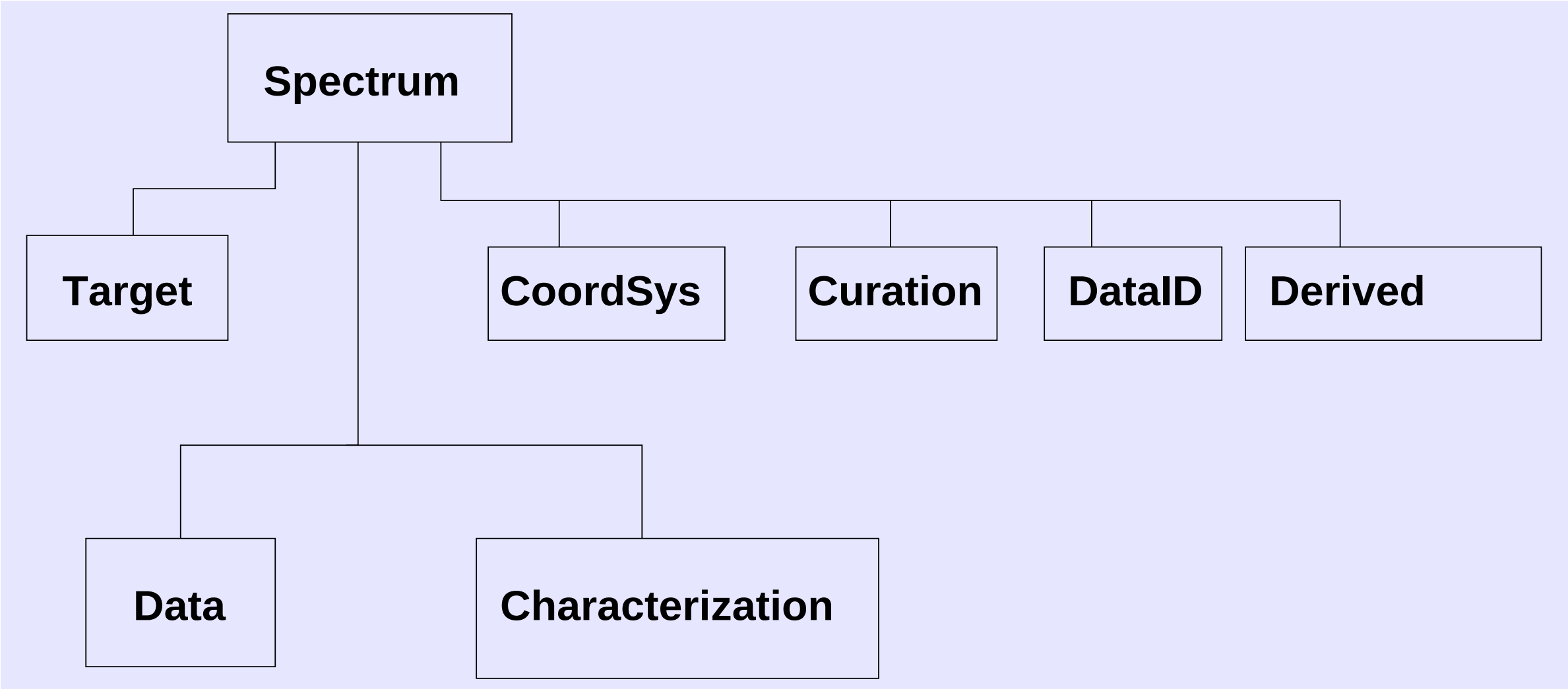

Figure 1: UML class diagram for the spectral data model.
The Characterization, Curation, DataID and Derived classes are shown in detail below in diagram form and with further text description in Section 5.

We specify fields that are MANDATORY (MUST), RECOMMENDED (SHOULD), or OPTIONAL (MAY). MANDATORY fields are in bold. MANDATORY means that a document **implementing the Spectrum use case** must provide a value; however, the value may be UNKNOWN (the value exists but is not known) or N/A (not applicable: for example, RA and DEC for a moving object or absolute time for a theory simulation). RECOMMENDED means that a data provider should try to fill the relevant fields if possible, but the document is still compliant if they are omitted. Particular serializations (FITS, VOTABLE, etc) may amend these requirements by specifying default values for the serialization.

The same data model can be reused by other IVOA standards in contexts where a spectrum is merely being **described**: notably, in the Simple Spectral Access Protocol where we are describing a **query about a spectrum**. **The list of mandatory, etc., fields does not apply to these other use cases.** The same data model fields and concepts, or a defined subset of them, should be used, but the list of mandatory, recommended and optional fields will be different and should be identified in the standards defining those use cases.

The minimal required content for the spectrum dataset use case is:

- Spectrum model version
- Target name and dataset title DataID.Title (which may be the same as each other)
- Characterization Coverage.Location and Coverage.Bounds (Extent or Start/Stop range) descriptions of the location and extent of the data in the RA, Dec, time and spectral domains
- The Curation.Publisher field
- The descriptions of the spectral coordinate and flux fields including UCD and units (Spectrum.Char.SpectralAxis, Spectrum.Char.FluxAxis)

- For each point: the values of the spectral coordinate and flux.

Note that each Spectrum instance has only one spectral coordinate axis. If you want to provide *both* flux-vs-wavelength and flux-vs-frequency for a single dataset, you must (in this version of the model) make two separate instances (VO resources).

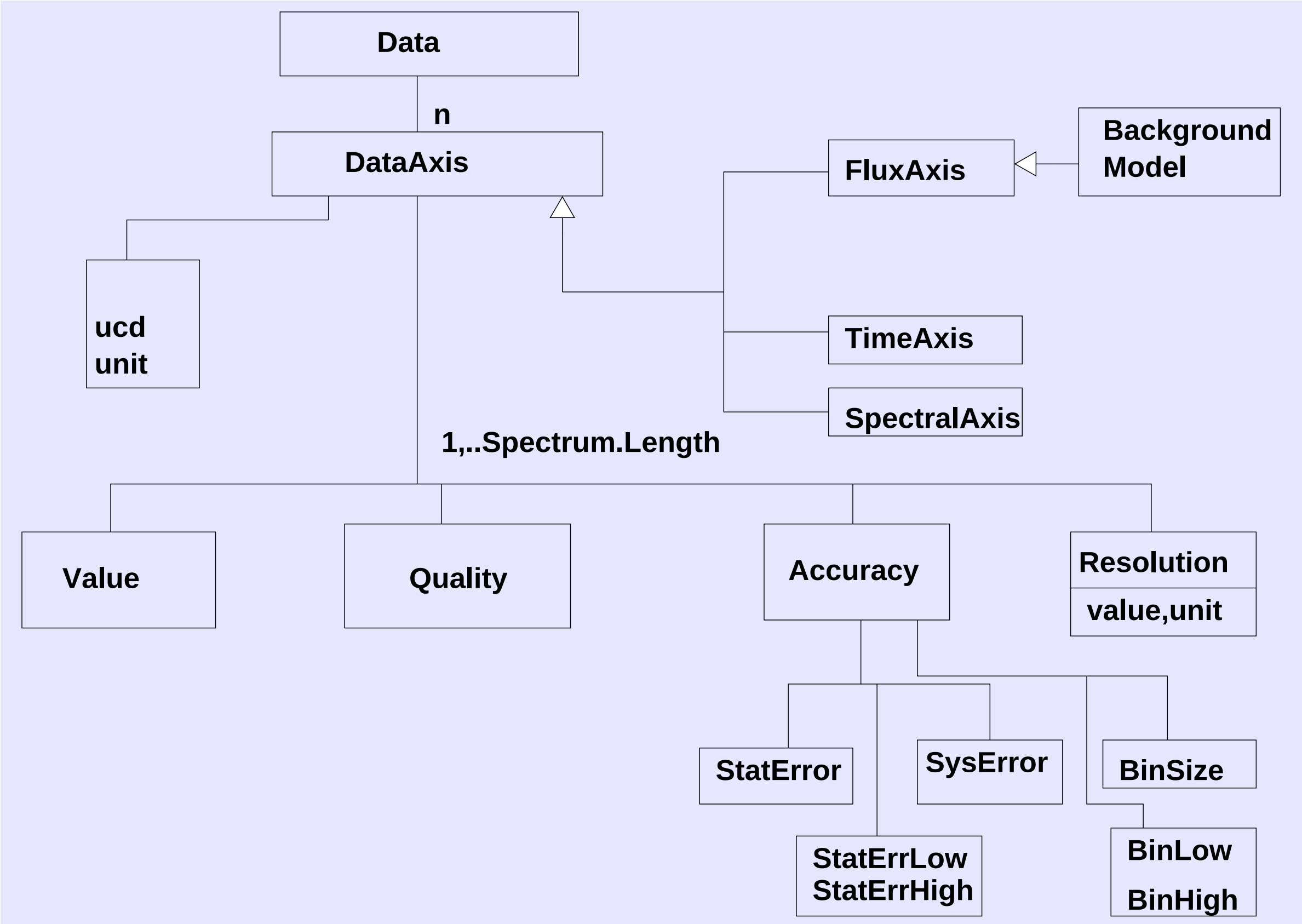

Figure 2: Diagram for Data object

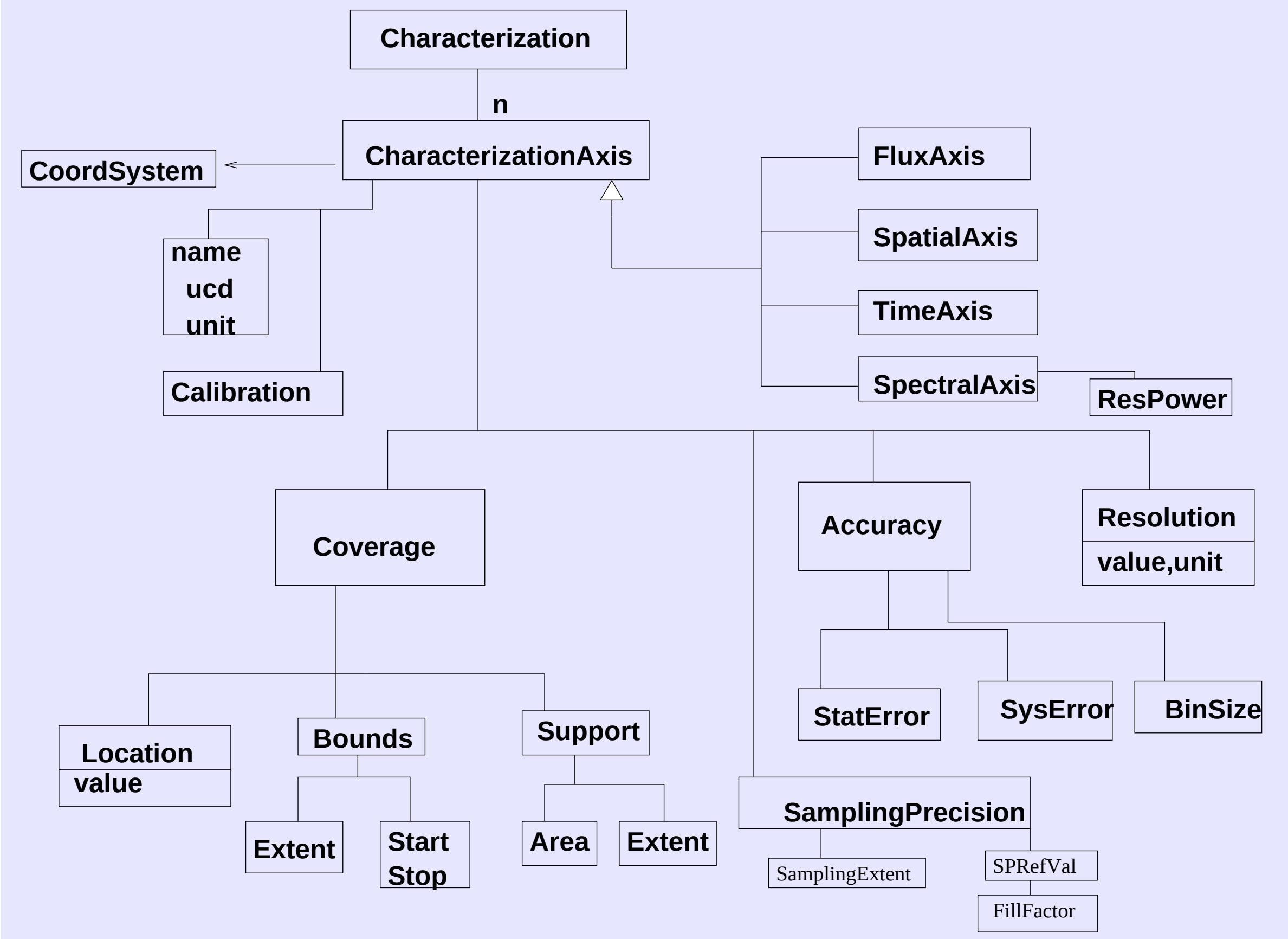

Figure 3: Diagram for Characterization object

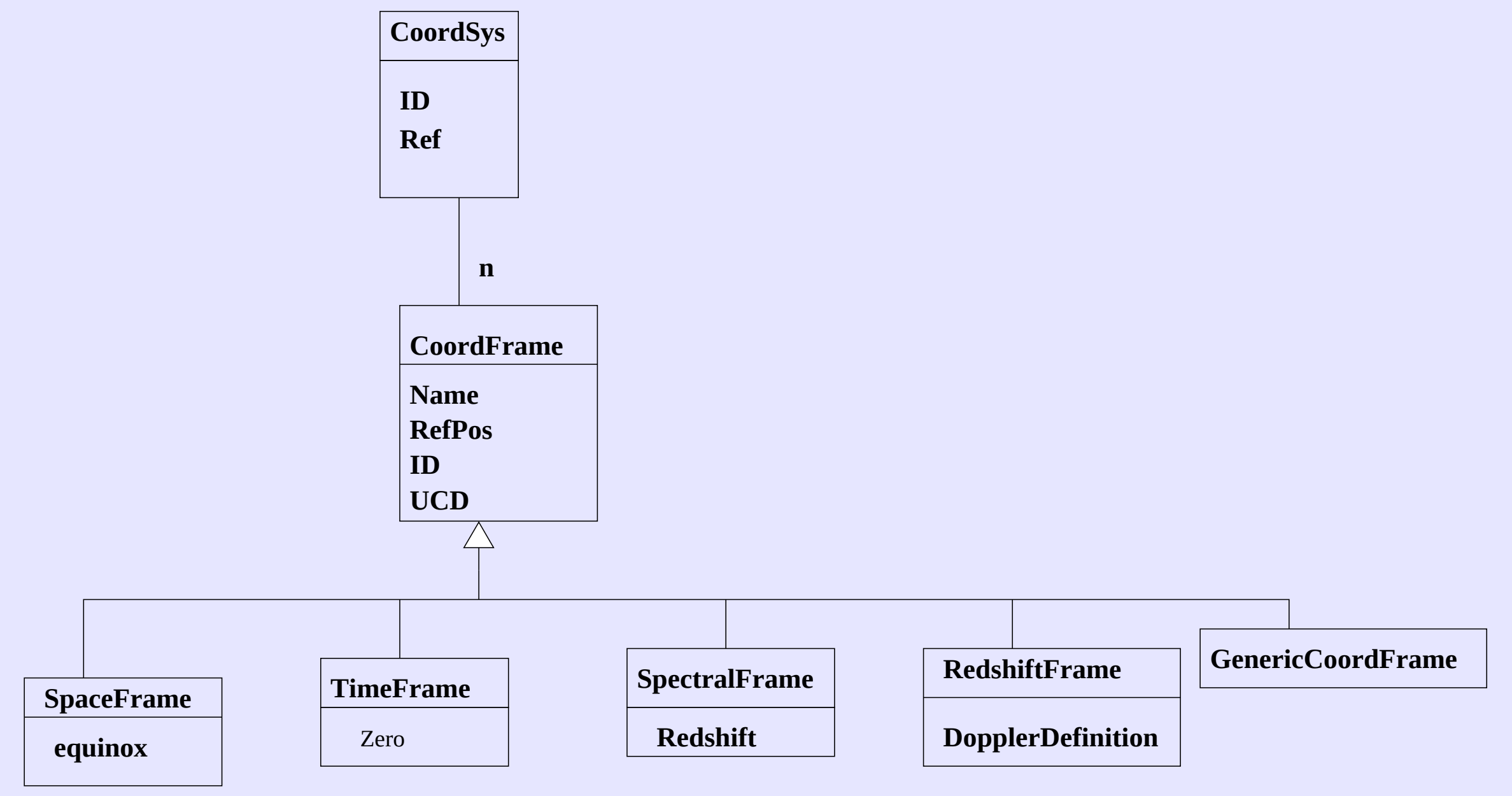

Figure 4: Diagram for CoordSys object

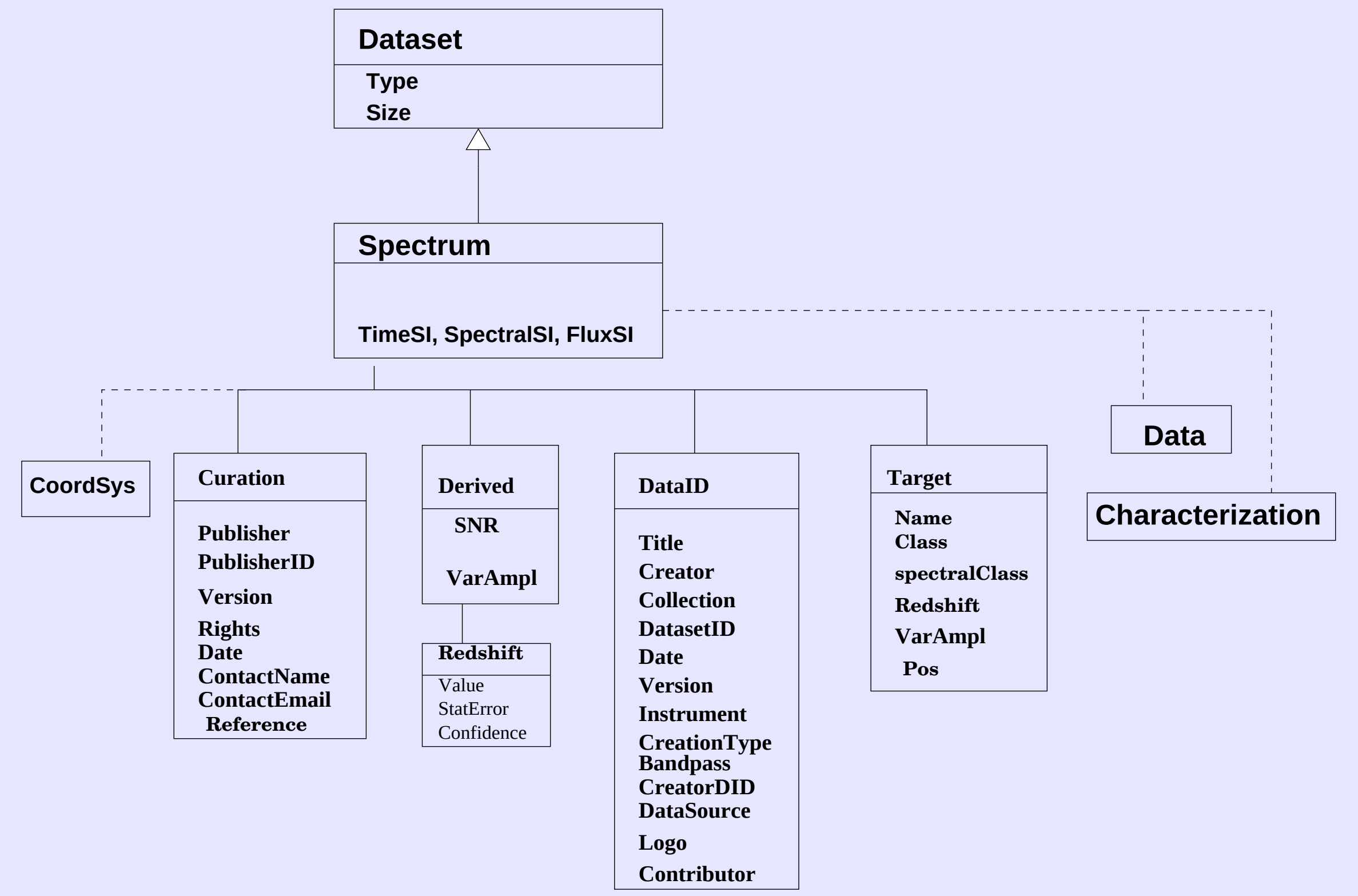

Figure 5: Diagram for remaining metadata: Curation, DataID, Derived, Target objects

### 3.3 Units

We adopt the WCS/OGIP convention for units: Document OGIP 93-001 (http://legacy.gsfc.nasa.gov/docs/heasarc/ofwg/docs/general/ogip_93_001/ogip_93_001.html).

Briefly, units are given in the form

```
10**(-14) erg/cm**2/s/Hz, 10**3 Jy Hz
```

i.e. with exponents denoted by **, division by /, multiplication by a space.

This format is mostly consistent with the AAS standards for online tables in journals (http://grumpy.as.arizona.edu/∼gschwarz/unitstandards.html) except for the use of space rather than "." for multiplication and the fact that we do not require the use of SI units.

SI prefixes for units are to be recognized; for instance, the listing of "m" as a known unit for wavelength implies that "cm", "nm", and "um" (with "u" the OGIP convention for rendering "micro") are also acceptable.

Until IVOA generic unit conversion software is mature and widely deployed, it is helpful to interoperable applications to include a representation of the units in "base SI form", including only the base units kg, m, s (and possibly A, sr) with a numeric prefix. Pedro Osuna and Jesus Salgado have proposed a representation in the spirit of dimensional analysis, using the symbols M, L, T to signify kg, m, s respectively and omitting the ** for powers, so that

```
10**3 Jy Hz
```

which is equivalent to

```
10**-23 kg s**-2
```

is written compactly as

```
1.E-23 MT-2
```

This alternate representation is supported for the main model fields (time, spectral coordinate and flux) only.

Although the spectral model is flexible enough to permit different units for each field, as a matter of style we strongly recommend that whenever possible the same units should be used for compatible fields (e.g. flux and error on flux).

### 3.4 UCDs

UCDs or Unified Content Descriptors are the IVOA's standardized vocabulary for astronomical concepts. In this document we use UCDs as field attributes (for example, element attributes in XML) to distinguish alternate physics within the same data model roles - for example, to distinguish frequency versus wavelength on the spectral coordinate 'X-axis'. In the list of UCDs below, the notation "em.*" is used to indicate "either em.wl, em.freq or em.energy"; exactly one of these must be used for any one instantiation of the model (so you can't put a literal em.* in a UTYPE field for a spectrum document, you should put whichever of em.wl etc. is appropriate for the actual data.)

The current list of valid UCDs is http://cdsweb.u-strasbg.fr/UCD/ucd1p-words.txt with syntax defined in the UCD recommendation http://www.ivoa.net/Documents/latest/UCD.html.

UCDs should be case insensitive.

### 3.5 UTYPEs and model reuse

UTYPE was a concept introduced in VOTABLE to label fields of a hierarchical data model. The word is now used generally to mean a standard identifier for a data model field. They are also case-insensitive and are of the form a.b.c.d where the dots indicate a 'has-a' hierarchy with the leftmost element being the containing model and the rightmost element being the lowest level element referred to. This is quite close to a simple XPATH in an XML schema, but we chose not to use slash instead of dot to emphasize that we are only specifiying the element type, not the exact position in an instance (so no sophisticated query syntax). We use the terms 'data model field' and 'UTYPE' interchangeably.

In the main text of this standard, the UTYPEs all begin with the prefix "Spectrum." to clarify their membership in the Spectrum data model. This is intended to reflect the main use case of a Spectrum dataset. Other IVOA standards (e.g. SSA) may use a different prefix instead of "Spectrum.", as long as they include a mechanism for unambigiously identifying the data model in use, e.g. by the value of the DataModel utype. This represents data model inheritance; we say that SSA inherits the Spectrum model, so that "SSA." utypes overlap with the Spectrum ones. In the main listing of UTYPEs in Table 1, the Spectrum prefix is omitted. Thus, when using Table 1 in conjunction with the Spectrum model (rather than reusing it in another model), UTYPEs such as 'Curation.Contact.Email' should be instantiated as 'Spectrum.Curation.Contact.Email'.

### 3.6 Packaging model

The simple Packaging model for SSA describes the format of the associated dataset. Allowed values for the format are briefly listed here; Detailed serialization for formats 4 to 6 are not specified; The metadata (format 7) is not returned by the standard SSA call; it instead uses a new getCapabilities option. See the SSA protocol definition document for details of this. These packaging values will be part of the SSA protocol response, and are implicit in the individual serializations. We only discuss formats 1 to 3 in this document.

- (1) FITS (standard BINTABLE for Spectrum, defined in this document)
- (2) VOTABLE
- (3) XML (native XML for web services and XML tools)
- (4) text (simple text table with columns of data and no markup)
- (5) text/html
- (6) graphics; a JPG, GIF etc. representation of the data
- (7) metadata; only the XML metadata.

### 3.7 Data Model Fields

The DM fields (or UTYPEs) for the Spectrum DM are tabulated on the following pages. The field names are to be used as the UTYPE values in VOTABLE serializations and in the TUTYPn keys in the FITS serialization.

The fields are explained in more detail in the following sections. They should have the prefix "Spectrum." when used in the spectrum dataset use case.

In Table 1, as well as the field names, a key to the FITS serialization is given. Where this key is blank, there is no FITS support for the field and it must take its default value. Several of the Char fields are required to be the same as the corresponding Data fields, and this is indicated by "(as Data)". Some fields describe properties of a FITS table column and use the values of keywords like TTYPEn, TUCDn for that column. The appropriate value of TTYPEn is indicated. In the case of columns related to the spectral coordinates, the appropriate TTYPEn value depends on the type of coordinate (WAVE for UCD em.wl, ENER for em.energy, FREQ for em.freq). This is indicated as e.g. "TTYPEn='WAVE_ELO',etc", meaning that the column name should be WAVE_ELO, FREQ_ELO, ENER_ELO as appropriate. The FITS serialization is described in more detail in a later section.

Table 1: Spectrum metadata fields

| Field | FITS | UCD1+ | Meaning | Req | Default |
|---|---|---|---|---|---|
| **DataModel** | VOCLASS | | Data model name and version | MAN | Spectrum-1.0 |
| Type | VOSEGT | - | Dataset or segment type | OPT | Spectrum |
| Length | DATALEN | meta.number | Number of points | OPT | (must be derived) |
| TimeSI | TIMESDIM | - | SI factor and dimensions | REC | |
| SpectralSI | SPECSDIM | - | SI factor and dimensions | REC | |
| FluxSI | FLUXSDIM | - | SI factor and dimensions | REC | |
| CoordSys.ID | VOCSID | | ID string for coordinate system | OPT | |
| CoordSys.SpaceFrame.Name | RADECSYS | | ICRS or FK5 | REC | ICRS |
| CoordSys.SpaceFrame.UCD | SKY_UCD | meta.ucd | Space frame UCD | OPT | Char.SpatialAxis.UCD |
| CoordSys.SpaceFrame.RefPos | SKY_REF | | Origin of SpaceFrame | OPT | UNKNOWN |
| CoordSys.SpaceFrame.Equinox | EQUINOX | time.equinox;pos.frame | Equinox | OPT | 2000.0 |
| CoordSys.TimeFrame.Name | TIMESYS | time.scale | Timescale | OPT | TT |
| CoordSys.TimeFrame.UCD | - | meta.ucd | Time frame UCD | OPT | time |
| CoordSys.TimeFrame.Zero | MJDREF | arith.zp;time | Zero point of timescale in MJD | OPT | 0.0 |
| CoordSys.TimeFrame.RefPos | - | time.scale | Times of photon arrival are at this location | OPT | TOPOCENTER |
| CoordSys.SpectralFrame.Name | SPECNAME | - | Spectral frame name | OPT | (None) |
| CoordSys.SpectralFrame.UCD | TUCDn | meta.ucd | Spectral frame UCD | OPT | Char.SpectralAxis.ucd |
| CoordSys.SpectralFrame.RefPos | SPECSYS | sdm:spect.frame | Spectral frame origin | OPT | TOPOCENTER |
| CoordSys.SpectralFrame.Redshift | REST_Z | | If restframe corrected | OPT | 0.0 |
| CoordSys.RedshiftFrame.Name | ZNAME | - | Redshift frame name | OPT | (None) |
| CoordSys.RedshiftFrame.DopplerDefinition | TCTYPnZ | - | Opt, Radio, or Rel. | OPT | UNKNOWN |
| CoordSys.RedshiftFrame.RefPos | SPECSYSZ | - | Redshift frame origin | OPT | UNKNOWN |
| **Curation.Publisher** | VOPUB | meta.curation | Publisher | MAN | |
| Curation.PublisherID | VOPUBID | meta.ref.url;meta.curation | URI for VO Publisher | OPT | UNKNOWN |
| Curation.Date | VODATE | | Date curated dataset last modified | OPT | UNKNOWN |
| Curation.Version | VOVER | meta.version;meta.curation | Version info | OPT | UNKNOWN |
| Curation.Rights | VORIGHTS | | Restrictions: public, proprietary, mixed | REC | Public |
| Curation.Reference | VOREF | meta.bib.bibcode | URL or Bibcode for documentation | REC | UNKNOWN |
| Curation.Contact.Name | CONTACT | meta.bib.author;meta.curation | Contact name | OPT | UNKNOWN |
| Curation.Contact.Email | EMAIL | meta.ref.url;meta.email | Contact email | OPT | UNKNOWN |
| Curation.PublisherDID | DS_IDPUB | meta.ref.url;meta.curation | Publisher's ID for the dataset ID | REC | DataID.DatasetID |

| Field | FITS | UCD1+ | Meaning | Req | Default |
|---|---|---|---|---|---|
| DataID.Title | TITLE | meta.title;meta.dataset | Dataset Title | MAN | None |
| DataID.Creator | AUTHOR | | VO Creator ID | OPT | UNKNOWN |
| DataID.Collection | COLLECTn | | Collection name(s) | OPT | None |
| DataID.DatasetID | DS_IDENT | meta.id;meta.dataset | IVOA Dataset ID | OPT | UNKNOWN |
| DataID.CreatorDID | CR_IDENT | meta.id | Creator's ID for the dataset | OPT | None |
| DataID.Date | DATE | time;meta.dataset | Data processing/creation date | OPT | UNKNOWN |
| DataID.Version | VERSION | meta.version;meta.dataset | Version of dataset | OPT | UNKNOWN |
| DataID.Instrument | INSTRUME | meta.id;instr | Instrument ID | OPT | UNKNOWN |
| DataID.Bandpass | SPECBAND | instr.bandpass | Band, consistent with RSM Coverage.Spectral | OPT | UNKNOWN |
| DataID.CreationType | CRETYPE | | dataset creation type (archive, cutout,derived) | OPT | Archival |
| DataID.Logo | VOLOGO | meta.ref.url | URL for creator logo | OPT | UNKNOWN |
| DataID.Contributor | CONTRIBn | | Contributor (may be many) | OPT | UNKNOWN |
| DataID.DataSource | DSSOURCE | | Original data type: survey, pointed, theory, artificial, composite | OPT | UNKNOWN |
| Derived.SNR | DER_SNR | stat.snr | Signal-to-noise for spectrum | OPT | UNKNOWN |
| Derived.Redshift.Value | DER_Z | | Measured redshift for spectrum | OPT | UNKNOWN (may be undefined) |
| Derived.Redshift.StatError | DER_ZERR | stat.error;src.redshift | Error on measured redshift | OPT | UNKNOWN |
| Derived.Redshift.Confidence | DER_ZCNF | | Confidence value on redshift | OPT | UNKNOWN |
| Derived.VarAmpl | DER_VAR | src.var.amplitude;arith.ratio | Variability amplitude as fraction of mean | OPT | UNKNOWN |
| **Target.Name** | OBJECT | meta.id;src | Target name | MAN | |
| Target.Description | OBJDESC | meta.note;src | Target descriptive text | OPT | UNKNOWN |
| Target.Class | SRCCLASS | src.class | Target or object class | OPT | UNKNOWN |
| Target.SpectralClass | SPECTYPE | src.spType | Object spectral class | OPT | UNKNOWN |
| Target.Redshift | REDSHIFT | src.redshift | Target redshift | OPT | UNKNOWN |
| Target.Pos | RA_TARG, DEC_TARG | pos.eq;src | Target RA and Dec | REC | UNKNOWN (may be variable) |
| Target.VarAmpl | TARGVAR | src.var.amplitude | Target variability amplitude, typical | OPT | UKNOWN |

| Field | FITS | UCD1+ | Meaning | Req | Default |
|---|---|---|---|---|---|
| | | Spectrum.Char Axis definition fields | | | |
| Char.FluxAxis.Name | (as Data) | | name for flux | OPT | Flux |
| **Char.FluxAxis.ucd** | (as Data) | meta.ucd | ucd for flux | MAN | |
| **Char.FluxAxis.unit** | (as Data) | meta.unit | Unit for flux | MAN | |
| Char.SpectralAxis.Name | (as Data) | | name for spectral axis | OPT | SpectralCoord |
| **Char.SpectralAxis.ucd** | (as Data) | meta.ucd | ucd for spectral coord | MAN | |
| **Char.SpectralAxis.unit** | (as Data) | meta.unit | Unit for spectral coord | MAN | |
| Char.TimeAxis.Name | - | | name for time axis | OPT | Time |
| Char.TimeAxis.ucd | - | meta.ucd | ucd for time | REC | time |
| Char.TimeAxis.unit | - | meta.unit | Unit for time | REC | d |
| Char.SpatialAxis.Name | - | meta.id | name for spatial axis | OPT | Sky |
| Char.SpatialAxis.ucd | SKY_UCD | meta.ucd | ucd for spectral coord | REC | pos.eq |
| Char.SpatialAxis.unit | - | meta.unit | Unit for spectral coord | REC | deg |
| Char.FluxAxis.Calibration | FLUX_CAL | | Type of coord calibration | OPT | CALIBRATED |
| Char.SpectralAxis.Calibration | SPEC_CAL | meta.code.qual | Type of coord calibration | OPT | CALIBRATED |
| Char.TimeAxis.Calibration | TIME_CAL | meta.code.qual | Type of coord calibration | OPT | CALIBRATED |
| Char.SpatialAxis.Calibration | SKY_CAL | meta.code.qual | Type of coord calibration | OPT | CALIBRATED |
| | | Spectrum.Char Coverage Fields | | | |
| **Char.SpatialAxis.Coverage.Location.Value** | RA,DEC | pos.eq | Position, usually ICRS | MAN | |
| **Char.SpatialAxis.Coverage.Bounds.Extent** | APERTURE | instr.fov | Aperture angular size, deg | MAN | |
| Char.SpatialAxis.Coverage.Support.Area | REGION | Aperture region | REC | UNKNOWN | |
| Char.SpatialAxis.Coverage.Support.Extent | AREA | instr.fov | Field of view area, sq deg | OPT | |
| **Char.TimeAxis.Coverage.Location.Value** | TMID | time.epoch | Exposure midpoint (MJD, d) | MAN | |
| **Char.TimeAxis.Coverage.Bounds.Extent** | TELAPSE | time.duration | Total elapsed time | MAN | |
| Char.TimeAxis.Coverage.Bounds.Start | TSTART | time.start;obs.exposure | Start time | REC | UNKNOWN |
| Char.TimeAxis.Coverage.Bounds.Stop | TSTOP | time.stop;obs.exposure | Stop time | REC | UNKNOWN |
| Char.TimeAxis.Coverage.Support.Extent | EXPOSURE | time.duration;obs.exposure | Effective exposure time | OPT | |
| **Char.SpectralAxis.Coverage.Location.Value** | SPEC_VAL | em.*;instr.bandpass | Spectral coord value | MAN | |
| **Char.SpectralAxis.Coverage.Bounds.Extent** | SPEC_BW | instr.bandwidth | Width of spectrum in A or other spec. coord. (See text) | MAN | |
| **Char.SpectralAxis.Coverage.Bounds.Start** | TDMINn | em.*;stat.min | Start in spectral coordinate | MAN | |
| **Char.SpectralAxis.Coverage.Bounds.Stop** | TDMAXn | em.*;stat.max | Stop in spectral coordinate | MAN | |
| Char.SpectralAxis.Coverage.Support.Extent | SPECWID | instr.bandwidth | Effective width of spectrum in A | OPT | |

| Field | FITS | UCD1+ | Meaning | Req | Default |
|---|---|---|---|---|---|
| | Spectrum.Char Sampling Fields | | | | |
| Char.SpectralAxis.SamplingPrecision. SampleExtent | SPEC_BIN | em.*;spect.binSize | Wavelength bin size | OPT | = Accuracy.BinSize |
| Char.SpatialAxis.SamplingPrecision. SampleExtent | TCDLTn | phys.angSize;instr.pixel | spatial bin size | OPT | Pixel size in deg |
| Char.TimeAxis.SamplingPrecision. SampleExtent | TIMEDEL | time.interval | time bin size | OPT | = Accuracy.BinSize |
| Char.SpatialAxis.SamplingPrecision. SamplingPrecisionRefVal.FillFactor | SKY_FILL | stat.filling;pos.eq | Sampling Filling factor | OPT | 1.0 |
| Char.SpectralAxis.SamplingPrecision. SamplingPrecisionRefVal.FillFactor | SPEC_FIL | stat.filling;em.* | Sampling Filling factor | OPT | 1.0 |
| Char.TimeAxis.SamplingPrecision. SamplingPrecisionRefVal.FillFactor | DTCOR | stat.filling;time | Sampling Filling factor | OPT | UNKNOWN |
| | Spectrum.Char Accuracy Fields - global | | | | |
| Char.FluxAxis.Accuracy.StatError | (as Data) | stat.error;phot.flux.density;em.* | error | REC | UNKNOWN |
| Char.FluxAxis.Accuracy.SysError | (as Data) | stat.error.sys;phot.flux.density;em.* | Systematic error | REC | UNKNOWN |
| Char.SpectralAxis.Accuracy.BinSize | SPEC_BIN | em.*;spect.binSize | Wavelength bin size | OPT (may be undefined) | UNKNOWN |
| Char.SpectralAxis.Accuracy.StatError | SPEC_ERR | stat.error;em.* | Spectral coord measurement error | REC | 0 |
| Char.SpectralAxis.Accuracy.SysError | SPEC_SYE | stat.error.sys;em.* | Spectral coord measurement error | REC | 0 |
| Char.SpectralAxis.Resolution | SPEC_RES | spect.resolution; em.* | Spectral resolution FWHM | OPT | Accuracy.BinSize |
| Char.SpectralAxis.ResPower | SPEC_RP | spect.resolution | Spectral resolving power | OPT | |
| Char.TimeAxis.Accuracy.BinSize | TIMEDEL | time.interval | Time bin size | OPT | UNKNOWN (undefined) |
| Char.TimeAxis.Accuracy.StatError | (as Data) | stat.error;time | Time coord measurement statistical error | OPT | UNKNOWN |
| Char.TimeAxis.Accuracy.SysError | (as Data) | stat.error.sys;time | Time coord measurement systematic error | OPT | UNKNOWN |
| Char.TimeAxis.Resolution | (as Data) | time.resolution | Temporal resolution FWHM | OPT | Accuracy.BinSize |
| Char.SpatialAxis.Accuracy.StatError | SKY_ERR | stat.error;pos.eq | Astrometric statistical error | OPT | UNKNOWN |
| Char.SpatialAxis.Accuracy.SysError | SKY_SYE | stat.error.sys;pos.eq | Systematic error | OPT | UNKNOWN |
| Char.SpatialAxis.Resolution | SKY_RES | pos.angResolution | Spatial resolution of data | OPT | UNKNOWN |

| Field | FITS | UCD1+ | Meaning | Req | Default |
|---|---|---|---|---|---|
| | | Data Fields | | | |
| **Data.FluxAxis.Value** | TTYPEn = 'FLUX' | (see sec 4.1, table 3) | Flux values for points | MAN | |
| Data.FluxAxis.ucd | TUCDn | meta.ucd | ucd for flux | OPT | Char.FluxAxis.ucd |
| Data.FluxAxis.unit | TUNITn | meta.unit | Unit for flux | OPT | Char.FluxAxis.unit |
| **Data.SpectralAxis.Value** | TTYPEn ='WAVE' or 'ENER' or 'FREQ' | (see sec 4.2, table 2) | Spectral coordinates for points | MAN | (Char.SpectralAxis.Location) |
| Data.SpectralAxis.ucd | TUCDn | meta.ucd | ucd for spectral coord | OPT | Char.SpectralAxis.ucd |
| Data.SpectralAxis.unit | TUNITn | meta.unit | Unit for spectral coord | OPT | Char.SpectralAxis.unit |
| Data.TimeAxis.Value | TTYPEn ='TIME' | | Time coordinates for points | OPT | Char.TimeAxis.Location |
| Data.TimeAxis.ucd | TUCDn | meta.ucd | ucd for time | OPT | Char.TimeAxis.ucd |
| Data.TimeAxis.unit | TUNITn | meta.unit | Unit for time | OPT | Char.TimeAxis.Unit |
| Data.BackgroundModel.Value | TTYPEn='BGFLUX' | | Flux values for points | OPT | No background model |
| Data.BackgroundModel.ucd | TUCDn | meta.ucd | ucd for background flux | OPT | Points.FluxAxis.ucd |
| Data.BackgroundModel.unit | TUNITn | meta.unit | Unit for background flux | OPT | Points.FluxAxis.unit |

| Field | FITS | UCD1+ | Meaning | Req | Default |
|---|---|---|---|---|---|
| Accuracy Fields - per data point (default to corresponding Spectrum.Char values) | | | | | |
| Data.FluxAxis.Accuracy.StatError | TTYPEn = 'ERR' | stat.error;phot.flux.density; em.* | symmetric error | OPT | (Char) |
| Data.FluxAxis.Accuracy.StatErrLow | TTYPEn='ERR_LO' | stat.error;phot.flux.density; em.*;stat.min | Lower error | OPT | (Char) |
| Data.FluxAxis.Accuracy.StatErrHigh | TTYPEn='ERR_HI' | stat.error;phot.flux.density; em.*;stat.max | Upper error | OPT | (Char) |
| Data.FluxAxis.Accuracy.SysError | TTYPEn='SYS_ERR' | stat.error.sys;phot.flux.density; em... | Systematic error | OPT | (Char) |
| Data.FluxAxis.Quality | TTYPEn='QUALITY' | meta.code.qual; phot.flux.density, em.* | Quality mask | OPT | 0 |
| Data.FluxAxis.Quality.n | TTYPEn='QUALn' | | String value, for n = 0,1,2..; meaning of quality value | OPT | None |
| Data.SpectralAxis.Accuracy.BinSize | TTYPEn='WAVE_BIN',etc | em.*;spect.binSize | Wavelength bin size | OPT | (Char) |
| Data.SpectralAxis.Accuracy.BinLow | TTYPEn ='WAVE_LO',etc | em.*;stat.min | Spectral coord bin lower end | OPT | |
| Data.SpectralAxis.Accuracy.BinHigh | TTYPEn ='WAVE_HI',etc | em.*;stat.max | Spectral coord bin upper end | OPT | |
| Data.SpectralAxis.Accuracy.StatError | TTYPEn ='WAVE_ERR',etc | stat.error;em.* | Spectral coord measurement error | OPT | (Char) |
| Data.SpectralAxis.Accuracy.StatErrLow | TTYPEn='WAVE_ELO',etc | stat.error; em.*;stat.min | Spectral coord measurement lower error | OPT | (Char) |
| Data.SpectralAxis.Accuracy.StatErrHigh | TTYPEn='WAVE_EHI',etc | stat.error; em.*;stat.max | Spectral coord measurement upper error | OPT | (Char) |
| Data.SpectralAxis.Accuracy.SysError | TTYPEn='WAVE_SYE',etc | stat.error.sys;em.* | Spectral coord systematic error | OPT | (Char) |
| Data.SpectralAxis.Resolution | TTYPEn ='WAVE_RES',etc | spect.resolution; em.* | Spectral resolution FWHM | OPT | (Char) |

| Field | FITS | UCD1+ | Meaning | Req | Default |
|---|---|---|---|---|---|
| Accuracy Fields - per data point (continued) | | | | | |
| Data.TimeAxis.Accuracy.BinSize | TIMEDEL | time.interval | Time bin size | OPT | (Char) |
| Data.TimeAxis.Accuracy.BinLow | TTYPEn='TIME_LO' | time.start;time.interval | Time bin start | OPT | |
| Data.TimeAxis.Accuracy.BinHigh | TTYPEn='TIME_HI' | time.end;time.inverval | Time bin stop | OPT | |
| Data.TimeAxis.Accuracy.StatError | TTYPEn='TIME_ERR' | stat.error;time | Time coord measurement statistical error | OPT | (Char) |
| Data.TimeAxis.Accuracy.StatErrLow | TTYPEn='TIME_ELO' | stat.error;time; stat.min | Time coord measurement lower error | OPT | (Char) |
| Data.TimeAxis.Accuracy.StatErrHigh | TTYPEn='TIME_EHI' | stat.error;time; stat.max | Time coord measurement upper error | OPT | (Char) |
| Data.TimeAxis.Accuracy.SysError | TTYPEn='TIME_SYE' | stat.error.sys;time | Time coord systematic error | OPT | (Char) |
| Data.TimeAxis.Resolution | TTYPEn='TIME_RES' | time.resolution | Temporal resolution FWHM | OPT | (Char) |
| Data.BackgroundModel.Accuracy. StatError | TTYPEn='BG_ERR' | stat.error;phot.flux.density; em.* | Symmetric error | OPT | (Char) |
| Data.BackgroundModel.Accuracy. StatErrLow | TTYPEn='BG_ELO' | stat.error;phot.flux.density; em.*;stat.min | Lower error | OPT | (Char) |
| Data.BackgroundModel.Accuracy. StatErrHigh | TTYPEn='BG_EHI' | stat.error;phot.flux.density; em.*;stat.max | Upper error | OPT | (Char) |
| Data.BackgroundModel.Accuracy. SysError | TTYPEn='BG_SYE' | stat.error.sys;phot.flux.density; em.* | Systematic error | OPT | (Char) |
| Data.BackgroundModel.Quality | TTYPEn ='BGQUAL' | meta.code.qual;phot.flux.density, em.* | Quality mask | OPT | 0 |

# 4 Spectral data model Measurement objects

## 4.1 Spectral coordinate

Astronomers use a number of different spectral coordinates to label the electromagnetic spectrum. The cases enumerated by Greisen et al. (2006) are listed below with their UCDs.

**MANDATORY: Exactly one Spectrum.Char.SpectralAxis field should be present, with units and one of the UCD values listed below.** We distinguish between the VO data model field name (which might be used for VOTABLE UTYPE), the FITS WCS name (provided for comparison only), and the UCD1+ names.

Note 1: For this version, only the first four entries, Wavelength, Frequency, Energy, and spectral channel, should be used for interoperable transmission of data - implementations are not required to understand (convert) the other UCD values.

Note 2: For the velocity cases, the UCD uses a spect.dopplerVeloc tree rather than a *src.veloc* tree, because the velocity here is really a labelling of a spectral coordinate, and the link to the physical radial velocity of the different emission sources contributing to the spectrum is rather indirect.

Note 3: 'em.wl;obs.atmos' (AWAV) is provided for air wavelengths. The basic spectral choices em.wl, em.freq, em.energy are understood to be vacuum values.

Table 2: Spectral coordinate options

| Field | FITS WCS | UCD1+ | Meaning | Units |
|---|---|---|---|---|
| PREFERRED CHOICES | | | | |
| SpectralAxis.ucd | **WAVE** | em.wl | Wavelength | ångstrom, m |
| SpectralAxis.ucd | **FREQ** | em.freq | Frequency of photon | Hz |
| SpectralAxis.ucd | **ENER** | em.energy | Photon energy | erg, eV, J |
| SpectralAxis.ucd | - | instr.pixel;em.wl | Instrumental spectral bin | chan |
| ALTERNATIVE CHOICES | | | | |
| SpectralAxis.ucd | **WAVN** | em.wavenumber | Wavenumber | m**(-1) |
| SpectralAxis.ucd | **AWAV** | em.wl;obs.atmos | Air wavelength | ångstrom, m |
| SpectralAxis.ucd | **WAVE-LOG** | em.wl | Log wavelength | |
| SpectralAxis.ucd | **FREQ-LOG** | em.freq | Log frequency of photon | |
| SpectralAxis.ucd | **ENER-LOG** | em.energy | Log photon energy | |
| SpectralAxis.ucd | **VELO** | spect.dopplerVeloc | Apparent radial velocity | m/s |
| SpectralAxis.ucd | **VRAD** | spect.dopplerVeloc.radio | Radio velocity | m/s |
| SpectralAxis.ucd | **VOPT** | spect.dopplerVeloc.opt | Optical velocity | m/s |
| SpectralAxis.ucd | **BETA** | spect.dopplerVeloc | Velocity (c=1) | - |

## 4.2 Flux (Spectral Intensity) Object

Two instances of the Flux object are supported: Flux and BackgroundModel. The Flux may be either the background-subtracted net flux or the total flux (the source+background), in the latter case hopefully with the BackgroundModel (see below). Net and total flux are distinguished by the 'src.net' UCD adjective.

For each of these cases, there are many slightly different physical quantities covered by the general concept of Flux; we distinguish them by their UCD. The table contains a list of flux quantities that applications should expect to read and handle. If you create a Spectrum instance with a flux quantity or flux unit not in the list below, you should expect that applications will be able to propagate it and recognize it, but not be able to merge it or compare it with other Spectrum instances. (For example, an application trying to measure line wavelengths shouldn't care too much that it doesn't understand what the flux units are).

Note in particular the distinction between the unit **count** (an instrumental value) and the unit **photon** (used in the photon number flux, i.e. the number of photons incident; photon number flux = energy flux divided by photon energy).

Note: The concept of the "nu L-nu" or "lambda L-lambda" luminosity flux, or equivalently the luminosity per logarithmic energy interval L(log nu), is a distinct concept in the world of spectral energy distributions - and it's a different concept from the bolometric luminosity, which has the same units. The UCD board has not yet approved a UCD expressing this concept; we have to use phys.luminosity and infer the concept from the units. My solution for brightness temperature is also rather questionable.

Note: we propose the UCD spect.continuum to represent continuum flux.

Table 3: Flux Value options

| Field | UCD1+ | Meaning | Unit (OGIP style) |
|---|---|---|---|
| FluxAxis.ucd | phot.flux.density;em.wl | Flux density per unit wave. | erg cm**(-2) s**(-1) angstrom**(-1), W m**(-2) m**(-1), keV cm**(-2) s**(-1) angstrom**(-1) |
| FluxAxis.ucd | phot.flux.density;em.freq | Flux density per unit freq. | erg cm**(-2) s**(-1) Hz**(-1), Jy, W m**(-2) Hz**(-1) |
| FluxAxis.ucd | phot.flux.density;em.energy | Flux density per energy interval | keV cm**(-2) s**(-1) kev**(-1) |
| FluxAxis.ucd | phot.flux.density;em.energy; meta.number | Photons per unit area, time, energy | photon cm**(-2) s**(-1) keV**(-1) |
| FluxAxis.ucd | phot.flux.density;em.wl | Flux density per log wave interval ($\nu F(\nu)$) | Jy Hz |
| FluxAxis.ucd | phot.flux.density.sb;em.wl | Surface brightness per unit wavelength | erg cm**(-2) s**(-1) angstrom**(-1) arcsec**(-2) |
| FluxAxis.ucd | phot.flux.density.sb;em.freq | Surface brightness per unit frequency | Jy sr**(-1) |
| FluxAxis.ucd | phot.count | Counts in spectral channel | count |
| FluxAxis.ucd | arith.rate;phot.count | Count rate in spectral channel | count/s |
| FluxAxis.ucd | arith.ratio;phot.flux.density | Flux ratio of two spectra | - |
| FluxAxis.ucd | phys.luminosity;em.wl | Luminosity per unit wave | erg s**(-1) angstrom**(-1), W/m |
| FluxAxis.ucd | phys.luminosity;em.freq | Luminosity per unit freq | erg s**(-1) Hz**(-1), W/Hz |
| FluxAxis.ucd | phys.luminosity;em.energy | Luminosity per unit energy | erg s**(-1) keV**(-1) |
| FluxAxis.ucd | phys.luminosity;em.energy | Luminosity per log frequency | erg s**(-1), W |
| FluxAxis.ucd | phys.energy.density | Radiation energy density per unit volume, per unit wave etc. | erg cm**(-3), J m**(-3) |
| FluxAxis.ucd | phot.fluence;em.wl | Photon number flux per unit wave. | photon cm**(-2) s**(-1) angstrom**(-1) |
| FluxAxis.ucd | phot.flux.density;em.wl; phys.polarization | Polarized flux per unit wavelength | erg cm**(-2) s**(-1) angstrom**(-1) |
| FluxAxis.ucd | phys.polarization | Polarized fraction vs spectral coord | (dimensionless) |
| FluxAxis.ucd | phys.luminosity; phys.angArea;em.wl | Flux per unit solid angle (at source) | erg cm**(-2) s**(-1) sr**(-1) angstrom**(-1) |
| FluxAxis.ucd | phot.antennaTemp | Antenna temperature | K |
| FluxAxis.ucd | phot.flux.density; phys.temperature | Brightness temperature | K |
| FluxAxis.ucd | phot.mag | Magnitude in defined band | mag |
| FluxAxis.ucd | phot.mag | AB (spectrophotometric) magnitude | mag |
| FluxAxis.ucd | phot.flux.density;instr.beam | Flux per resolution element (e.g. Jy/beam) | Jy/beam |
| FluxAxis.ucd | phot.mag.sb | Surface brightness in magnitudes | mag arcsec**(-2) |
| FluxAxis.ucd | phys.transmission | Filter transmission, 0.0 to 1.0 | (dimensionless) |
| FluxAxis.ucd | phys.area;phys.transmission | Effective area | cm**2 |
| FluxAxis.ucd | phot.flux.density;em.wl; spect.continuum | Continuum only | erg cm**(-2) s**(-1) angstrom**(-1) arcsec**(-2) |

### 4.3 BackgroundModel Object

We optionally allow a BackgroundModel value for each Flux value. We define NetFlux = TotalFlux - BackgroundModel. The name BackgroundModel, rather than Background, reminds us that it is an estimate: often, the BackgroundModel will be generated by taking a flux measurement at another location and rescaling it for any difference in exposure time or extraction aperture.

The BackgroundModel array is required to have the same UCD and units as the Flux array. It represents a model for the expected flux values if the Target had zero flux.

**OPTIONAL: There may be at most one BackgroundModel.Value field present. It must have the same UCD as the Flux.**

### 4.4 Time coordinate

For data with a time-series component, whether regularly sampled or sparse photometry points, the time coordinate is given by an elapsed time in some physical units (e.g. seconds or days) relative to a reference time.

This reference time is given in MJD as the field Spectrum.Char.TimeAxis.Coverage.Location, as described in the Characterization section. For a simple spectrum with no time-resolved data, this is the time of the observation (ideally the midpoint).

For time-resolved data, the time coordinate Spectrum.Data.TimeAxis.Value refers to the midpoint of the sample interval. See the Space-Time Coordinates document for details of time coordinate complexity.

The time unit is specified by a string, and the only valid values for this unit are 's' (seconds) and 'd' (days).

### 4.5 Position coordinate

In general we may consider position coordinates as part of the measurement (and possibly varying from point to point), but this capability is not included in the current document. The (celestial) position of the aperture for the spectrum is given in the spatial Spectrum.Char.SpatialAxis.Coverage.Location field. The Spectrum.Char.SpatialAxis.Coverage.Location.Value field is in the coordinates of CoordSys.SpaceFrame. The default is ICRS RA,Dec in decimal degrees.

### 4.6 Accuracy Fields

We include accuracy models for both the coordinates (spectral, spatial and temporal) and the fluxes. The accuracy can appear in two places: in the global characterization, where it represents typical accuracy for the dataset, and in the data points themselves, providing a way to provide per-data-point errors. All the Accuracy fields are optional, both in the per-data-point fields and in the Characterization instances; the per-data fields default to the values in Characterization.

#### 4.6.1 Coordinate bins

We express the bandpass for each spectral bin as a low and high value for the spectral coordinate, or as a width. The same is done for photometry points, which amounts to approximating a filter by a rectangular bandpass. Time bins are also given as low and high values or as a width. Note that the width values are suitable for Spectrum.Char (the global accuracy) while the bin low/high values only have meaning for Spectrum.Data (the per-data-point values).

Only one of BinSize, or both BinLow and BinHigh, must be present (possibly as a header parameter implying a constant value for each flux point). If absent, the bin limits are assumed to be halfway between the coordinate values and bounded by the range given in Char.*.Coverage.Extent.

#### 4.6.2 Uncertainties

In addition to the binning, we allow the model to express uncertainties (which may be larger than the bin width), both statistical and systematic. We allow one or two-sided statistical errors but only one-sided systematic errors. You can specify StatErr, or StatErrHigh/StatErrLow, but not both. Statistical errors which have the same units as the data, and systematic errors which are dimensionless fractions (e.g. a 5 percent systematic error is expressed as 0.05).

For position we have a single statistical error - a two-sided error doesn't make sense for a 2D coordinate. Eventually we may want a full error ellipse, but this is too complicated for the present model.

We also use a very simple error model for the fluxes: we include plus and minus flux errors, and a quality flag. The errors are understood as 1 sigma gaussian errors which are uncorrelated for different points in the spectrum. If the data provider has only upper limit information, it should be represented by setting the flux value and the lower error value equal to the limit, and the upper error value equal to zero (e.g. 5 (+0,-5)). In general applications may choose to render measurements as upper limits if the flux value is less than some multiple (e.g. 3) of the lower error. We also allow a systematic error value, assumed constant across a given spectrum and fully correlated (so that, e.g. it does not enter into estimating spectral slopes).

**CLARIFICATION:** the two-sided errors StatErrLow and StatErrHigh are the plus/minus ERRORS, not the (value+error, value-error). In other words, if Value = 10 and there is a symmetric uncertainty of 3, the ErrorLow and ErrorHigh are both +3.0, and NOT 7.0, 13.0. This is different from the sampling description BinLow and BinHigh, which give the VALUES at the low and high end of the bin. Thus if the central wavelength of the bin is 4200.0, and the bin size is 10, then the BinLow and BinHigh values are 4195.0, 4205.0 and NOT 10.0, 10.0. Note that because of this, 0.0 is NOT an acceptable default for BinLow and BinHigh, while it IS acceptable (albeit unlikely) for StatErrLow and StatErrHigh.

The StatErrLow, StatErrHigh, SysError fields for SpectralCoord, Time, Sky and Flux are optional; however, omitting these fields indicates that the errors are unknown. Data providers are STRONGLY encouraged to provide explicit error measures whenever possible.

#### 4.6.3 Resolution

We also include a trivial resolution model: a single number nominally representing a FWHM spectral or time resolution expressed in the same units as the spectral or time coordinate. The default is to assume that the resolution is equal to the BinSize if defined. The spatial (sky) resolution may be useful to know if it exceeds the aperture size; the default is to assume it is equal to the aperture size.

For the spectral characterization, we allow an alternative field called the spectral resolving power: Spectrum.Char.SpectralAxis.ResPower: this is the dimensionless $\lambda/\Delta\lambda$. It is often preferred for spectra because it is often more constant across the spectrum than the resolution. ResPower and Resolution can be interchanged by dividing out Coverage.Location.

Similar quantities can't really be defined for temporal and spatial resolving power since there's no absolute time or spatial scale, so we call the spectral one out as a special case. One could define a temporal or spatial frequency using the bounds - i.e. just the number of resolution elements in the spectrum - but that's a slightly different concept.

#### 4.6.4 Quality

The Quality model represents quality by an integer, with the following meanings: 0 is good data, 1 is data which is bad for an unspecified reason (e.g., no data in the sample interval), and other positive integers greater than 1 may be used to flag data which is bad or dubious for specific reasons.

The data provider may also define scalar string-valued metadata fields Quality.2, Quality.3... to define specific quality flags on a per-spectrum basis. Bitmasks, used in some archives such as SDSS, should be remapped to such independent Quality fields.

Quality defaults to zero, i.e. good data.

#### 4.6.5 Calibration

We also introduce a Calibration field which can have the values ABSOLUTE, RELATIVE, NORMALIZED or UNCALIBRATED. This is expected to be particularly useful to describe the flux.

ABSOLUTE indicates that the values in the data are expected to be correct within the given uncertainty.

RELATIVE indicates that although an unknown systematic error is present, the ratio of any two values will be correct.

NORMALIZED indicates that the data have been divided by a reference spectrum; the flux UCD in this case should be suffixed by 'arith.ratio;' and the units should be blank (dimensionless).

UNCALIBRATED indicates that although the values reflect a measurement of the given UCD, they are modified by an unspecified coordinate-dependent correction. Such values may be useful in the case of a spectrum with ABSOLUTE calibration on the wavelengths but UNCALIBRATED fluxes; the wavelengths of discontinuous features such as spectral lines can be measured on the assumption that the missing calibration function has no sharp discontinuities in the region of interest.

The Calibration fields are present in the CharacterizationAxis elements.

# 5 Associated Metadata Fields

Most of the associated metadata are generic observational metadata that can be applied to future data models, and are not specific to spectra.

## 5.1 CoordSys Fields

The CoordSys object is a simplified instance of the STC CoordSystem object. For XML serializations, it can be replaced by an actual STC CoordSystem instance.

CoordSys consists of 1 or more CoordFrame objects, each of which defines the coordinates for a particular axis. The CoordSys has an overall ID string, which is user-defined and arbitrary. Each CoordFrame also has a type, a UCD and a ReferencePosition; the Reference Position gives the origin of the coordinate system (and thus also its rest frame).

For the space, time, and spectral axes we define specialized CoordFrames for convenience: SpaceFrame, TimeFrame and SpectralFrame. The CoordFrame names (types) for SpaceFrame and TimeFrame must be from a controlled list; for other frames, the type is an arbitrary string.

Note: For compatibility with the Characterization schema, data model elements Spectrum.Char.SpatialAxis.CoordSys, etc. are allowed, but in Spectrum these must be trivial references to the overall Spectrum.CoordSys.

| Token | Meaning | Note |
|---|---|---|
| UNKNOWN | Unknown origin | |
| RELOCATABLE | Relative origin | Suitable for simulations |
| CUSTOM | Origin specified wrt another system | |
| TOPOCENTER | Location of the observing device | (telescope) |
| | | |
| BARYCENTER | Solar system barycenter | |
| HELIOCENTER | Center of the Sun | |
| GEOCENTER | Center of the Earth | |
| | | |
| EMBARYCENTER | Earth-Moon barycenter | |
| MOON | Center of the Moon | |
| MERCURY | Center of Mercury | |
| VENUS | Center of Venus | |
| MARS | Center of Mars | |
| JUPITER | Center of Jupiter | |
| SATURN | Center of Saturn | |
| URANUS | Center of Uranus | |
| NEPTUNE | Center of Neptune | |
| PLUTO | Center of Pluto | |
| | | |
| LSRK | Kinematic local standard of rest | Redshift frame only |
| LSRD | Dynamic local standard of rest | Redshift frame only |
| GALACTIC_CENTER | Center of the Galaxy | |
| LOCAL_GROUP_CENTER | Barycenter of the Local Group | |

Table 4: Allowed values for CoordFrame.ReferencePosition

### 5.1.1 SpaceFrame

The SpaceFrame has an optional Equinox attribute which is used if the frame name is FK4 or FK5. The allowed frame names for SpaceFrame are listed below.

### 5.1.2 TimeFrame

The TimeFrame is defined by the frame name and the ReferencePosition. Allowed values of the name are given below.

One standard reference time in astronomy is the origin of Julian Day Number on the TT (Terrestrial Time) timescale, BC 4713 Nov 24 at 11:59:27.81 (Gregorian). Using TT is preferable to UTC because it does not contain leap seconds, so the elapsed time in days is just equal to the difference in JD values.

The ISO-8601 calendar format standard does not support dates before AD 1, so cannot express this reference time. Therefore, it is not a suitable format for internal representations of such reference times. However, non-default choices of reference time may be specified in external serializations by a date in ISO-8601 format, e.g. "2004-11-30T11:59:00.01".

In this version of the model we require use of MJD as the time type for absolute times. (ISO dates and JD are other possibilities covered by the STC document). Relative times in a time series may be in other units, relative to the TimeFrame.Zero value.

(Note that in the FITS serialization, the MJDREF keyword allows definition of reference times in decimal days relative to MJD 0.0 = JD 2400000.5.)

### 5.1.3 SpectralFrame

The spectral frame is defined by its ReferencePosition. Once the choice of wavelength versus frequency or energy has been made, the only free parameter is the location at which the spectrum would have the given spectral coordinates. For directly observed data this is the topocenter (location of the observation); spectra may be velocity-corrected to a given velocity frame, which may be defined by the location which is at rest in that velocity frame (e.g. the heliocenter). Strictly, the correction may not be just a velocity shift, but any kind of spectral shift including e.g. gravitational redshifts; it is still true that such a shift corresponds to a location (e.g. surface of a white dwarf star) that can be quoted as a reference position.

Since the frame is defined by its ReferencePosition, the frame name is not important, and will not be significant to software. We suggest that it may be filled by the name of the spectral coordinate, using FITS names such as 'WAVE', 'FREQ' or 'ENER'.

The spectral frame has an optional Redshift attribute to specify a rest frame; it is used only if the frame's ReferencePosition is "CUSTOM". This redshift is measured in dimensionless units, defined as $\Delta\lambda/\lambda$ and may be negative. No specific interpretation of the shift as a cosmological or velocity shift effect is implied; we note for the record that some co-authors object to using the word 'redshift' in this generic sense.

### 5.1.4 RedshiftFrame (also the Velocity frame)

When you convert the spectral coordinate to velocity or redshift (relative to some assumed rest-frame spectral feature) you need to record some other metadata. Our field name containing this metadata is RedshiftFrame, but we emphasize that the name redshift does not imply that blueshifts are excluded, merely that, in both galactic and extragalactic astronomy, when a shift is interpreted as a velocity a positive value indicates a shift to the red. The concept of Redshift frame includes both cosmological and local Doppler velocities.

Note that you only use RedshiftFrame if you're measuring things in velocities; a rest-frame spectrum of a redshifted quasar whose spectral axis is in ångstroms will be described by a SpectralFrame. The reason we have BOTH SpectralFrame and RedshiftFrame is to support certain data products, particularly used in spectral line radioastronomy, in which a spectrum (possibly obtained in piecewise spectral regions) is refactored into a set of separate spectral segments centered on different spectral lines; each segment is assigned a velocity axis centered on that line (and the same pixel from the original spectrum can appear in multiple segments each with a different velocity coordinate); you then consider the data as a 2D array with a spectral axis (indexing the segments) and a velocity axis (for each segment/spectral line).

Other coordinate system information needed for velocity spectral coordinates include the observation-fixed spectral frame, the observatory location, the source redshift, and the velocity zero point (in Greisen et al, SSYSOBS, OBSGEO, VELOSYS, RESTFRQ/RESTWAV). However, we omit these in the current model. The only metadata we provide is the Doppler Definition - optical, radio or pseudo-relativistic.

#### 5.1.5 STC

**Notes on compatibility with, and differences from, STC 1.0:**

- We add the extra Redshift attribute in the SpectralFrame, instead of the more complex CustomReferencePosition approach used in STC.
- In STC's XML serialization, the frame types and reference positions are enumerated defined elements. Here they are strings, and we require that the frame name is the frame type.
- We don't explicitly include the coordinate flavor.

**OPTIONAL: All CoordSys values are optional** , but data providers should take special care to check whether or not the defaults are appropriate for their data. The implications of the defaults are:

- Positions are given in ICRS RA,Dec in degrees and are heliocentric values (i.e. corrected for annual parallax and aberration, as normally found in source catalogs).
- Times are given in MJD days and represent times of photon arrival at the telescope.
- Spectral coordinates are as observed at the telescope, and not corrected for redshift, the motion of the Earth, etc.

| Token | Meaning | Parameter(s) |
|---|---|---|
| UNKNOWN | Unknown frame | |
| CUSTOM | Custom frame | Pole, axis |
| AZ_EL | Azimuth and elevation | |
| BODY | Generic body (eg planet) | |
| | | |
| ICRS | The ICRS frame | |
| FK4 | FK4 | Equinox |
| FK5 | FK5 | Equinox |
| ECLIPTIC | Ecliptic l,b | Equinox |
| GALACTIC_I | Old galactic LI,BI | |
| GALACTIC_II | Galactic LII,BII | |
| SUPER_GALACTIC | SGL, SGB | |
| | | |
| MAG | Geomagnetic ref frame | |
| GSE | Geocentric Solar Ecliptic | |
| GSM | Geocentric Solar Magnetic | |
| SM | Solar Magnetic | |
| HGC | Heliographic | |
| HEE | Heliocentric Earth Ecliptic | |
| HEEQ | Heliocentric Earth Equatorial | |
| HCI | Heliocentric Inertial | |
| HCD | Heliocentric of Date | |
| | | |
| GEO_C | Geocentric corotating | |
| GEO_D | Geodetic ref frame | Spheroid |
| MERCURY_C | Corotating planetocentric | |
| VENUS_C | Corotating planetocentric | |
| LUNA_C | Corotating planetocentric | |
| MARS_C | Corotating planetocentric | |
| JUPITER_C_III | Corotating planetocentric | |
| SATURN_C_III | Corotating planetocentric | |
| URANUS_C_III | Corotating planetocentric | |
| NEPTUNE_C_III | Corotating planetocentric | |
| PLUTO_C | Corotating planetocentric | |
| MERCURY_G | Corotating planetographic | |
| VENUS_G | Corotating planetographic | |
| LUNA_G | Corotating planetographic | |
| MARS_G | Corotating planetographic | |
| JUPITER_G_III | Corotating planetographic | |
| SATURN_G_III | Corotating planetographic | |
| URANUS_G_III | Corotating planetographic | |
| NEPTUNE_G_III | Corotating planetographic | |
| PLUTO_G | Corotating planetographic | |

Table 5: Allowed values for CoordSys.SpaceFrame.Name

| Token | Meaning | Note |
|---|---|---|
| LOCAL | Relocatable (simulation) time | |
| TT | Terrestrial Time | |
| UTC | Coordinated Universal Time | |
| ET | Ephemeris Time | |
| TDB | Barycentric dynamical time | |
| TCG | Terrestrial Coordinate Time | |
| TCB | Barycentric Coordinate Time | |
| TAI | International Atomic Time | |
| LST | Local Sidereal Time | |

Table 6: Allowed values for CoordSys.TimeFrame.Name

### 5.2 Characterization

The Characterization metadata in this document are consistent with the IVOA Characterization data model draft as of March 2007. The Characterization model has a set of CharacterizationAxis objects. Each CharacterizationAxis describes the axis, and contains a Coverage describing the scope of the data, and optionally a Resolution and a Sampling object.

The CharacterizationAxis is identified by its UCD attribute. Spectrum instances should have Spatial, Time and Spectral characterization axes as well as FluxAxis.

To simplify things for the common axes, we define SpatialAxis, SpectralAxis, TimeAxis objects as special cases of CharacterizationAxis.

The CoordSystem element in CharacterizationAxis is there for compatibility with the Characterization document and, if present, should be a simple reference to the main Spectrum CoordSystem.

The Characterization fields will have a constant value for a given spectrum.

Note: In the SSA protocol/query response, we will restrict the Char units to meters (spectral coord), seconds (time coord), and decimal degrees (spatial), for simplicity and consistency with other parameters. We allow a more general approach for the full Spectrum instance (returned serializations); the units may be as described elsewhere in this document.

#### 5.2.1 Coverage Fields

The coverage fields will have a constant value for a given spectrum. They describe the region of space, time and spectrum from which the data were taken. In the Characterization model, we define progressively more accurate descriptions of this region: Location gives a single characteristic point, Bounds gives a range within which the data lies, and Support gives the detailed spatial field of view footprint, on/off time ranges (including gaps) and spectral ranges. (A fourth level not yet supported, Sensitivity, will provide detailed depth information: exposure map, time sensitivity variation, spectral transmission curve).

There is a field for giving the effective exposure time (useful for selecting among multiple spectra from the same instrument). The aperture field is important to determine what part of an extended object is contributing to the spectrum; we allow a simple aperture description (Char.SpatialAxis.Coverage.Bounds.Extent) consisting of a single number representing the aperture size in decimal degrees. For a slit spectrum, the effective aperture on the sky is usually the slit width in the cross-dispersion direction, while for a fiber it may be a circular region. For an accurate description, a full region polygon is allowed in the Area field. Note that since the goal of the VO Spectrum description is to describe the data as it is now, not to describe where it came from, our 'aperture' is always the effective extraction aperture, not the original instrument aperture if that is different.

The units of the spectral Coverage.Bounds.Extent (or Coverage.Bounds.Start/Stop) and Coverage.Support should be the same as those of SpectralCoord.

For time, the Coverage.Bounds.Start/Stop is a pair of values giving the start and stop time. Coverage.Bounds.Extent is the total elapsed time (Stop - Start) while Coverage.Support.Extent is the effective exposure time (total length of all observing intervals times any statistical deadtime filling factor). In the full Characterization model, Coverage.Support provides a whole array of start-stop pairs indicating data accumulated over a series of intervals. We may add this to the Spectrum model in a later revision.

The SpatialAxis.Coverage.Location and SpatialAxis.Coverage.Bounds.Extent, TimeAxis.Coverage.Location are required, as are either TimeAxis.Coverage.Bounds.Extent or TimeAxis.Coverage.Bounds.Start and Stop. If Extent is provided, Start and Stop are defined to be (Location - 0.5* Extent, Location +0.5*Extent).

The spectral equivalents, SpectralAxis.Coverage.Location and SpectralAxis.Bounds.Start/Stop, are also required in the model; serializations may decide to omit them since they are easily derived from the data.

The SamplingPrecision.SamplingPrecisionRefVal.FillFactor (previously Coverage.Support.Fill) fields give the filling factor, a statistical way of indicating that an axis is only partly sampled. The full IVOA Characterization data model provides a more detailed SamplingPrecision tree; although we fill only part of this we retain the field names for compatibility.

FillFactor is used for dead time corrections (time axis), statistical corrections for gaps between active pixels (spatial axis), and so on. Its value should be between 0 and 1, with the default being 1. (Although we provide a SpectralAxis FillFactor for symmetry and completeness, we are not aware of any practical application for it).

### 5.2.2 Region definitions

In the optional Char.SpatialAxis.Coverage.Support.Area we describe the detailed aperture shape in absolute coords on the sky. However, we don't allow a full STC region description. Our simplified region model allows for (1) a circle and (2) a polygon in a string representation: either

circle x0 y0 r

or

polygon x1 y1 x2 y2 x3 y3 ...

for example

```
circle 233.70 -13.32 0.00043
polygon 233.70 -13.32 233.71 -13.30 ...
```

where the positions and radii are required to be in degrees, in the coordinate system defined by CoordSys.

## 5.3 Derived Data Fields

The Derived (short for Derived Data) object has useful, and optional, summary information about the spectrum. For now, we include the option of adding signal-to-noise and variability indicators and a measurement of the redshift.

### 5.3.1 Signal-to-noise

The signal-to-noise is provided mainly as a way for searches to exclude data whose quality is insufficient for a particular study. Data providers may use their own definition, as we do not prescribe a uniform method to calculate it. A suitable method, set forth by the STScI/ST-ECF/CADC Spectral Container Working Group, is to define the signal as the median of the flux values in the spectrum and the noise as the median absolute third-order difference of flux values spaced two pixels apart. This noise value is then multiplied by 1.482602 / sqrt(6). Padded zeros in the flux values are skipped. A detailed description and discussion of the algorithm can be found in the issue #42 of the ST-ECF newsletter[5]. Implementations of the algorithm can be obtained from the ST-ECF website[6].

This method describes the high-spectral-frequency noise but does not take into account intermediate-spectral-frequency background 'noise'; projects which are background dominated may wish to include this in the noise definition. Furthermore most spectra vary in SNR across their waveband; users should therefore only use this single SNR as a crude selection parameter.

### 5.3.2 Redshift measurement model

One common piece of derived data for a spectrum is the source redshift. We provide fields for both the redshift measured value and statistical error. As above, we define the redshift to be $\Delta\lambda/\lambda$ and it may be positive or negative. The Derived field represents a measurement of the redshift from the data; a field in the Target object is available to store the redshift of the source as known from other means.

We add a further optional measure of accuracy, the Confidence, which expresses a probability between 0 and 1 that the quoted errors do apply. This measure is used in the Sloan spectral service to provide a way of describing the estimated probability that the redshift is completely in error because the lines have been misidentified. Its default value is 1.0.

In general, such a Confidence could be useful for any measurement where the error probability distribution has multiple peaks in parameter space, and could later be added to the standard Accuracy model.

Note that there are two other redshifts in our model: the Target redshift, a useful piece of metadata particularly for extragalactic objects, considered as an externally known property of the target (and so defined even if no lines are visible in the spectrum); and the SpectralFrame redshift, used only if a "rest frame" spectrum is presented and representing the assumed redshift used to shift the spectrum.

### 5.3.3 Variability amplitude

The variability amplitude field allows data providers to supply a characteristic amplitude (a precise value is not required). It is dimensionless; a value of 0.2 implies a 20 percent variation around the mean value.

## 5.4 Curation model

The Curation is an object consistent with the Curation information in the document "Resource Metadata for the Virtual Observatory Version 1.01", although some of the fields from RM curation have been moved to the DataID object, as discussed in the SSAP protocol document.

In Curation, we have added a Reference field for a bibliographic or documentation reference (this can occur multiple times), Rights field (same as Resource.Rights) for public/proprietary, and PublisherDID for a publisher-specified IVORN to the data. The Curation.PublisherDID is the same as the Resource Metadata V1.10 Resource.Identifier.

Version is provided by the publisher or creator and may be any string.

Curation.Publisher is REQUIRED. All other fields are optional.

[5] http://www.spacetelescope.org/about/further_information/newsletters/html/newsletter_42.html

[6] http://www.stecf.org/software/ASTROsoft/DER_SNR/

### 5.5 Data Identification model

The Data Identification model gives the dataset ID for a particular spectrum, and its membership of larger collections. All DataId fields are optional.

There are three dataset idenfifiers in the model: one under Curation and two here. All of them comply with the description of dataset identifiers as specified by the IVOA[7], including the use of 'stop' characters to identify specific datasets that are not individually in the registry.

The DataID.CreatorDID is the dataset ID defined internally by the creator and may be entirely different from the DatasetID described above. It is used to identify a particular original exposure in an archive and will not necessarily change even if the VO object in question is a cutout or is otherwise further processed.

The Curation.PublisherDID is a dataset ID defined by a publisher of the data. It may be an internal ID used by the archive.

The DataID.DatasetID may be the same as Curation.PublisherDID; for this field we recommend a journal-based URI such as the IVOA/ADEC/ADS dataset identifier. By agreement between the AAS journals, the ADS and the ADEC (NASA data centers), dataset identifiers[8] will be used to link journal articles back to the archival datasets containing the relevant observational data. If analogous but independent systems of URI designation are later adopted by other centers (e.g. by European journals) and accepted by IVOA, they will be suitable in this field.

For example, a dataset held by an archive which curates many missions and telescopes may have an ID allocated by the original mission (CreatorDID), an ID used as an index by the multi-mission archive (PublisherDID), and the ADS-style ID (DatasetID). These may all be different, although we hope that many archives will choose to use the ADS ID as their index.

We introduce the concept of a dataset creation type, which can have one of the following values, described in more detail in the SSAP protocol document.

- Archival, indicating that it is one of a collection of datasets (in this case spectra) generated in a systematic, homogeneous way and stored statically (or at least versioned). It will be possible to regenerate this dataset at a later date. The remaining types imply on-the-fly manipulation.
- Cutout, indicating that the dataset was created 'on-the-fly', by subsetting, but not by modifying values.
- Filtered, which may involve excluding data prior to binning into samples, also on the fly
- Mosaic, combining multiple original datasets on the fly
- spectral extraction, e.g. from a spectral data cube;
- catalog extraction [Not relevant for Spectrum model].

The dataset is associated with one or more Collections (instrument name, survey name. etc.) indicating some degree of compatibility with other datasets sharing the same Collection properties. Examples of possible Collection values are: "WFC", "Sloan", "BFS Spectrograph", "MSX Galactic Plane Survey".

[7] http://www.ivoa.net/Documents/latest/IDs.html

[8] http://vo.ads.harvard.edu/dv/

We also include a DataID.Bandpass, which is a string describing the spectral range. It can be one of the strings in Resource-Service-Metadata's Spectral.Coverage (e.g. "Optical") or Spectral.Coverage.Bandpass (e.g. "B" ). At the moment there is no fixed list of values for the RSM Spectral.Coverage.Bandpass.

For DataSource, see the SSAP protocol document.

### 5.6 Target model

In spectral data it is particularly important to be able to specify the target of the observation, which may be an astronomical source or some other target (calibration, diffuse background, etc.). By explicitly including a target model we can not only facilitate searches on particular types of target but also support archives of model spectra for which the Coverage fields may not be relevant. The Target.Name field is required; all other Target fields are optional.

The Target.pos field gives a nominal RA and Dec for the target, for example the catalog position of the source; the Coverage.Location fields in the spectrum indicate the actual telescope pointing position for that spectrum. (An SED might have a single Target object with a known position, but many Spectrum objects with slightly different telescope pointings). Similarly, the Target.redshift is the assumed actual redshift of the astronomical object, if applicable (again, usually from a catalog, NED, etc.), while the redshifts in the Derived objects in the spectrum (segment) indicates a redshift measured from that spectrum. The Target.redshift is normally used to store the cosmological redshift of extragalactic objects, although it may also be used to store the observed redshift of Galactic sources if that information is felt by the data provider to be useful.

At the moment there is no international standard list of valid values for Target class and spectral class. Nevertheless an initial deployment of the VO would gain some benefit from using archive-specific classes, and provide a framework for converging on a standard list.

### 5.7 Spectrum top level object

The Spectrum object contains the Data object with the actual data; the Target and Derived objects; and the standard dataset metadata of CoordSys, Characterization, Curation and DataID. We also add a CustomParams field to allow for propagation of unmodelled application-specific metadata.

In addition, we add an SIDim field for each axis giving the SI units of the values in the Osuna-Salgado dimensional format.

In spectral associations (such as SED applications), the Spectrum model is reused for both Spectrum and TimeSeries and is renamed Segment. The Spectrum object is expected to be generalized to a higher level Dataset object.

Each Spectrum (or Segment) may have a Length attribute giving the number of flux points in the data (in some serializations this value is deduced from the size of the data arrays, while in others it is made explicit).

Each Spectrum (or Segment) may also have a Type attribute indicating whether the data is intended as a TimeSeries (data are same spectral coord, varying times), Photometry (data are different spectral coords with irregular gaps), Spectrum (data are different spectral coords in contiguous bins), or Mixed (some mixture of the above).

This attribute is optional and defaults to Spectrum.

Segments are discussed in more detail in the Spectral Associations document which describes SEDs and other groupings.

# 6 Relationship to general VO data models

The Spectrum model involves objects addressed by the proposed VO Observation and Quantity data models. Although these models have not yet been fully worked out, we may note that a single Spectrum maps to the Observation model, which will include the Curation and Characterization objects. The Flux and the spectral coordinate entries together with their associated errors and quality will be special cases of the Quantity model, as will the simpler individual parameters. The field structure presented here is consistent with current drafts of the models.

## 6.1 Extensibility

The model and serializations defined in this document are extensible in the following sense:

- Future versions of the abstract (UML) model may add attributes or fields, and may deprecate the 'optional' property of existing fields.
- The Characterization object may have extra 'generic' axes, but these are not considered to be part of the Spectrum model. See Characterization specification for more.
- For the FITS serialization, implementors may add arbitrary additional keywords or table columns; readers must be able to handle files containing extra keywords and columns, and are encouraged to propagate such extra information when copying files. This permits local conventions to be layered on the basic definition.
- For the VOTABLE serialization, implementors may add arbitrary additional GROUP, PARAM or FIELDref/FIELD elements, with the restriction that the layering of existing elements should not be changed. (e.g. within the spectrum:DataID GROUP one may add a new GROUP containing newly defined PARAMs, but one may not move the existing Title PARAM inside the new group because that would change its indentation level). Readers must be able to handle files containing extra elements and are encouraged to propagate such extra information when copying files. This permits local conventions to be layered on the basic definition.
- For the XML object-based serialization, the CustomParams element at the top level of Spectrum is intended to allow extensibility and is equivalent to the ability to add a new GROUP at the top level. Future versions of the schema could use type extension to back-compatibly include the current schema as a special case, but apart from the CustomParams we have not currently provided for local extensibility within the current schema. (We could improve the schema by allowing the Group element to have arbitrary extra Param elements.)

# References

McDowell, Tody 2011, IVOA Simple Spectral Access Protocol V1.1
http://www.ivoa.net/Documents/SSA/

Martinez, Derriere 2007, The UCD1+ Controlled Vocabulary Version 1.23
http://www.ivoa.net/Documents/cover/UCDlist-20070402.html

Louys et al. 2008, Data Model for Astronomical DataSet Characterisation
http://www.ivoa.net/Documents/latest/CharacterisationDM.html

Rots 2007, Space-Time Coordinate Metadata for the Virtual Observatory
http://www.ivoa.net/Documents/latest/STC.html

Ochsenbein et al. 2004, VOTable Format Definition V1.1
http://www.ivoa.net/Documents/REC/VOTable/VOTable-20040811.html

Greisen, EW, Valdes F G, Calabretta M R and Allen S L 2006, A&A 446, 747.

Hanisch, R., (ed)., Resource Metadata for the VO, Version 1.01, 2004 Apr 26.
http://www.ivoa.net/Documents/latest/RM.html

Derriere, S. et al (eds.), UCD, Moving to UCD 1+, 2004 Oct 26.
http://www.ivoa.net/Documents/latest/UCD.html

# Part 2: XML schema serialization

# 7 XML schema serialization

## 7.1 XML schema

In the following XML schema, we implement the model fairly directly.

Within a spectrum the data points are kept together in objects called Point. Also, we have included a CustomParams element to allow site-specific metadata to be added.

The Coverage.Location fields have been collapsed to simple values rather than SEDCoord elements; this should perhaps be extended in a future version.

The Flux object is defined as an example of a more general SEDQuantity object, which is also used for the Sloan spectral service's redshift information.

A SED aggregation model is also included in the schema, as the top level element. This may be ignored until the SED model has been approved by IVOA.

```
<?xml version="1.0" encoding="utf-8"?>
<xs:schema xmlns="http://www.ivoa.net/xml/Spectrum/Spectrum-1.01.xsd"
xmlns:xs="http://www.w3.org/2001/XMLSchema"
xmlns:jxb="http://java.sun.com/xml/ns/jaxb"
xmlns:xlink="http://www.w3.org/1999/xlink"
targetNamespace="http://www.ivoa.net/xml/Spectrum/Spectrum-1.01.xsd"
elementFormDefault="qualified" jxb:version="1.0">
<xs:import namespace="http://www.w3.org/1999/xlink" schemaLocation="http://www.ivoa.net/xml/Xlink/xlink.xsd"/>

<!-- Customization for code generation with JAXB: not required otherwise -->
<xs:annotation>
  <xs:appinfo>
    <jxb:globalBindings
        generateIsSetMethod="true"/>
  </xs:appinfo>
 </xs:annotation>

<!-- A single segment corresponding to a spectrum or single point -->

<xs:element name="BaseSegment" type="segmentType"/>
<xs:element name="Spectrum" type="spectrumType" substitutionGroup="BaseSegment"/>
<xs:element name="Segment" type="spectrumType" substitutionGroup="BaseSegment"/>
<xs:element name="TimeSeries" type="timeSeriesType" substitutionGroup="BaseSegment"/>

<xs:complexType name="spectrumType">
<xs:complexContent mixed="false">
<xs:extension base="segmentType"/>
</xs:complexContent>
</xs:complexType>
<xs:complexType name="timeSeriesType">
<xs:complexContent mixed="false">
<xs:extension base="segmentType"/>
</xs:complexContent>
</xs:complexType>

<xs:complexType name="segmentType">
<xs:complexContent mixed="false">
<xs:extension base="Group">
<xs:sequence>
<xs:element minOccurs="0" maxOccurs="1" name="Target" type="targetType" />
<xs:element minOccurs="0" maxOccurs="1" name="Char" type="characterizationType" />
<xs:element minOccurs="0" maxOccurs="1" name="CoordSys" type="coordSysType" />
<xs:element minOccurs="0" maxOccurs="1" name="Curation" type="curationType" />
<xs:element minOccurs="0" maxOccurs="1" name="DataID" type="dataIDType" />
<xs:element minOccurs="0" maxOccurs="1" name="Derived" type="derivedDataType" />
<xs:element minOccurs="0" maxOccurs="1" name="CustomParams" type="arrayOfParamType" />
<xs:element minOccurs="0" maxOccurs="1" name="Type" type="textParamType" />
<xs:element minOccurs="0" maxOccurs="1" name="Length" type="intParamType" />
<xs:element minOccurs="0" maxOccurs="1" name="TimeSI" type="textParamType" />
<xs:element minOccurs="0" maxOccurs="1" name="SpectralSI" type="textParamType" />
<xs:element minOccurs="0" maxOccurs="1" name="FluxSI" type="textParamType" />
<xs:element minOccurs="0" maxOccurs="1" ref="Data"/>
</xs:sequence>
</xs:extension>
</xs:complexContent>
</xs:complexType>
```

```
<!-- The top level element: an SED with one target and many segments -->
<xs:element name="SED" nillable="true" type="sedType" />

<xs:complexType name="sedType">
<xs:sequence>
<xs:element minOccurs="0" maxOccurs="1" name="Date" type="timeParamType" />
<xs:element minOccurs="0" maxOccurs="1" name="Target" type="targetType" />
<xs:element minOccurs="0" maxOccurs="1" name="CustomParams" type="arrayOfParamType" />
<xs:element minOccurs="0" maxOccurs="1" name="Type" type="textParamType" />
<xs:element minOccurs="0" maxOccurs="1" name="NSegments" type="intParamType" />
<xs:element minOccurs="0" maxOccurs="unbounded" ref="BaseSegment"/>
<xs:element minOccurs="0" maxOccurs="1" name="Creator" type="textParamType" />
<xs:element minOccurs="0" maxOccurs="1" name="CreatorDID" type="textParamType" />
<xs:element minOccurs="0" maxOccurs="1" name="SpectralMinWavelength" type="doubleParamType" />
<xs:element minOccurs="0" maxOccurs="1" name="SpectralMaxWavelength" type="doubleParamType" />
</xs:sequence>
</xs:complexType>


<!-- Define the UCDs etc for the SED coordinate and the flux coordinate,
and include a global to specify accuracy etc which happens to be
constant for the entire segment (note that in SEDCoord,
value has minOccurs=0 so it can be omitted) -->
<!-- A single SEDCoord (time or spectral coord) value, or two values if it is spatial. -->

<xs:complexType name="sedBaseCoordType">
<xs:complexContent mixed="false">
<xs:extension base="Group"/>
</xs:complexContent>
</xs:complexType>

<xs:complexType name="sedCoordType">
<xs:complexContent mixed="false">
<xs:extension base="sedBaseCoordType">
<xs:sequence>
<xs:element minOccurs="0" maxOccurs="1" name="Value" type="doubleParamType" />
<xs:element minOccurs="0" maxOccurs="1" name="Accuracy" type="accuracyType" />
<xs:element minOccurs="0" maxOccurs="1" name="Resolution" type="doubleParamType" />
</xs:sequence>
</xs:extension>
</xs:complexContent>
</xs:complexType>

<xs:complexType name="sedQuantityType">
<xs:complexContent mixed="false">
<xs:extension base="sedBaseCoordType">
<xs:sequence>
<xs:element minOccurs="0" maxOccurs="1" name="Value" type="doubleParamType" />
<xs:element minOccurs="0" maxOccurs="1" name="Accuracy" type="accuracyType" />
<xs:element minOccurs="0" maxOccurs="1" name="Resolution" type="doubleParamType" />
<xs:element minOccurs="0" maxOccurs="1" name="Quality" type="intParamType" />
</xs:sequence>
</xs:extension>
</xs:complexContent>
</xs:complexType>
```

```
<!-- A set of useful types to add UCDs and units to base types; like BasicQuantity -->
<xs:complexType name="Group" >
<xs:attribute name="id" type="xs:ID" use="optional"/>
<xs:attribute name="idref" type="xs:IDREF" use="optional"/>
</xs:complexType>


<xs:complexType name="textParamType">
<xs:simpleContent>
<xs:extension base="paramType" />
</xs:simpleContent>
</xs:complexType>

<xs:complexType name="paramType">
<xs:simpleContent>
<xs:extension base="xs:string">
<xs:attribute name="name" type="xs:string" />
<xs:attribute name="ucd" type="xs:string" />
</xs:extension>
</xs:simpleContent>
</xs:complexType>

<xs:complexType name="dateParamType">
<xs:simpleContent>
<xs:extension base="paramType" />
</xs:simpleContent>
</xs:complexType>


<xs:complexType name="positionParamType">
<xs:sequence>
<xs:element minOccurs="2" maxOccurs="2" name="value" type="doubleParamType" />
</xs:sequence>
</xs:complexType>

<xs:complexType name="doubleParamType">
<xs:simpleContent>
<xs:extension base="paramType">
<xs:attribute name="unit" type="xs:string" />
</xs:extension>
</xs:simpleContent>
</xs:complexType>

<xs:complexType name="timeParamType">
<xs:simpleContent>
<xs:extension base="paramType">
<xs:attribute name="unit" type="xs:string" />
</xs:extension>
</xs:simpleContent>
</xs:complexType>

<xs:complexType name="intParamType">
<xs:simpleContent>
<xs:extension base="paramType">
<xs:attribute name="unit" type="xs:string" />
</xs:extension>
</xs:simpleContent>
</xs:complexType>
```

```
<!-- The error model. Bin entries will usually be omitted for the flux coordinate -->
<xs:complexType name="accuracyType">
<xs:complexContent mixed="false">
<xs:extension base="Group">
<xs:sequence>
<xs:element minOccurs="0" maxOccurs="1" name="BinLow" type="doubleParamType" />
<xs:element minOccurs="0" maxOccurs="1" name="BinHigh" type="doubleParamType" />
<xs:element minOccurs="0" maxOccurs="1" name="BinSize" type="doubleParamType" />
<xs:element minOccurs="0" maxOccurs="1" name="StatError" type="doubleParamType" />
<xs:element minOccurs="0" maxOccurs="1" name="StatErrLow" type="doubleParamType" />
<xs:element minOccurs="0" maxOccurs="1" name="StatErrHigh" type="doubleParamType" />
<xs:element minOccurs="0" maxOccurs="1" name="SysError" type="doubleParamType" />
<xs:element minOccurs="0" maxOccurs="1" name="Confidence" type="doubleParamType" />
</xs:sequence>
</xs:extension>
</xs:complexContent>
</xs:complexType>


<!-- The Field type allows us to define what our axes are -->
<xs:complexType name="arrayOfFieldType">
<xs:sequence>
<xs:element minOccurs="0" maxOccurs="unbounded" name="Field" nillable="true" type="fieldType" />
</xs:sequence>
</xs:complexType>

<xs:complexType name="fieldType">
<xs:attribute name="name" type="xs:string" />
<xs:attribute name="unit" type="xs:string" />
<xs:attribute name="ucd" type="xs:string" />
</xs:complexType>

<!-- The Point type groups a single set of time, spectral and flux values -->
<xs:element name="Data" type="arrayOfGenPointType"/>

<xs:complexType name="arrayOfGenPointType"/>

<xs:complexType name="arrayOfPointType">
<xs:complexContent mixed="false">
<xs:extension base="arrayOfGenPointType">
<xs:sequence>
<xs:element minOccurs="0" maxOccurs="unbounded" name="Point" nillable="true" type="pointType" />
</xs:sequence>
</xs:extension>
</xs:complexContent>
</xs:complexType>

<xs:element name="ArrayOfPoint" type="arrayOfPointType" substitutionGroup="Data"/>

<xs:complexType name="pointType">
<xs:sequence>
<xs:element name="TimeAxis" minOccurs="0" maxOccurs="1" type="sedCoordType" />
<xs:element name="SpectralAxis" minOccurs="0" maxOccurs="1" type="sedCoordType" />
<xs:element name="FluxAxis" minOccurs="0" maxOccurs="1" type="sedQuantityType" />
<xs:element name="BackgroundModel" minOccurs="0" maxOccurs="1" type="sedQuantityType" />
</xs:sequence>
</xs:complexType>
```

```xml
<xs:element name="ArrayOfFlatPoint" type="arrayOfFlatPointType" substitutionGroup="Data"/>

<xs:complexType name="arrayOfFlatPointType">
<xs:complexContent mixed="false">
<xs:extension base="arrayOfGenPointType">
<xs:sequence>
<xs:element minOccurs="0" maxOccurs="unbounded" name="Point" nillable="true" type="flatPointType" />
</xs:sequence>
</xs:extension>
</xs:complexContent>
</xs:complexType>



<xs:complexType name="flatPointType">

<xs:attribute name="T" type="xs:double"/>
<xs:attribute name="T_BinL" type="xs:double" />
<xs:attribute name="T_BinH" type="xs:double" />
<xs:attribute name="T_Size" type="xs:double" />
<xs:attribute name="T_Res" type="xs:double" />
<xs:attribute name="SP" type="xs:double" />
<xs:attribute name="SP_BinL" type="xs:double" />
<xs:attribute name="SP_BinH" type="xs:double" />
<xs:attribute name="SP_Size" type="xs:double" />
<xs:attribute name="SP_Res" type="xs:double" />
<xs:attribute name="F" type="xs:double" />
<xs:attribute name="F_ErrL" type="xs:double" />
<xs:attribute name="F_ErrH" type="xs:double" />
<xs:attribute name="F_Sys" type="xs:double" />
<xs:attribute name="F_Qual" type="xs:int" />
<xs:attribute name="BG" type="xs:double" />
<xs:attribute name="BG_ErrL" type="xs:double" />
<xs:attribute name="BG_ErrH" type="xs:double" />
<xs:attribute name="BG_Sys" type="xs:double" />
<xs:attribute name="BG_Qual" type="xs:int" />
</xs:complexType>
```

```
<!-- Now we define the higher level metadata -->
<xs:complexType name="contactType">
<xs:complexContent mixed="false">
<xs:extension base="Group">
<xs:sequence>
<xs:element minOccurs="0" maxOccurs="1" name="Name" type="textParamType" />
<xs:element minOccurs="0" maxOccurs="1" name="Email" type="textParamType" />
</xs:sequence>
</xs:extension>
</xs:complexContent>
</xs:complexType>

<xs:complexType name="curationType">
<xs:complexContent mixed="false">
<xs:extension base="Group">
<xs:sequence>
<xs:element minOccurs="0" maxOccurs="1" name="Publisher" type="textParamType" />
<xs:element minOccurs="0" maxOccurs="1" name="PublisherID" type="textParamType" />
<xs:element minOccurs="0" maxOccurs="1" name="Reference" type="textParamType" />
<xs:element minOccurs="0" maxOccurs="1" name="Version" type="textParamType" />
<xs:element minOccurs="0" maxOccurs="1" name="Contact" type="contactType" />
<xs:element minOccurs="0" maxOccurs="1" name="Rights" type="textParamType" />
<xs:element minOccurs="0" maxOccurs="1" name="Date" type="dateParamType" />
<xs:element minOccurs="0" maxOccurs="1" name="PublisherDID" type="textParamType" />
</xs:sequence>
</xs:extension>
</xs:complexContent>
</xs:complexType>


<xs:complexType name="characterizationType">
<xs:complexContent mixed="false">
<xs:extension base="Group">
<xs:sequence>
<xs:element name="SpatialAxis" type="characterizationAxisType" minOccurs="0" maxOccurs="1"/>
<xs:element name="TimeAxis" type="characterizationAxisType"  minOccurs="0" maxOccurs="1"/>
<xs:element name="SpectralAxis" type="spectralCharacterizationAxisType"  minOccurs="0" maxOccurs="1"/>
<xs:element name="FluxAxis" type="characterizationAxisType"  minOccurs="0" maxOccurs="1"/>
<xs:element minOccurs="0" maxOccurs="unbounded" name="CharacterizationAxis"
                              type="characterizationAxisType"/>
</xs:sequence>
</xs:extension>
</xs:complexContent>
</xs:complexType>
```

```
<xs:complexType name="characterizationAxisType">
<xs:complexContent mixed="false">
 <xs:extension base="Group">
 <xs:sequence>
 <xs:element minOccurs="0" maxOccurs="1" name="CoordSystem" type="coordSysType" />
 <xs:element minOccurs="0" maxOccurs="1" name="Coverage" type="coverageType" />
 <xs:element minOccurs="0" maxOccurs="1" name="Resolution" type="doubleParamType"/>
 <xs:element minOccurs="0" maxOccurs="1" name="Accuracy" type="accuracyType" />
 <xs:element minOccurs="0" maxOccurs="1" name="SamplingPrecision" type="samplingPrecisionType" />
 <xs:element minOccurs="0" maxOccurs="1" name="Calibration" type="textParamType" />
 </xs:sequence>
 <xs:attribute name="name" type="xs:string" />
 <xs:attribute name="ucd" type="xs:string" />
 <xs:attribute name="unit" type="xs:string" />
 </xs:extension>
 </xs:complexContent>
</xs:complexType>



<xs:element name="CharacterizationAxis" type="characterizationAxisType"/>

  <xs:complexType name="coverageType">
    <xs:complexContent mixed="false">
      <xs:extension base="Group">
        <xs:sequence>
          <xs:element minOccurs="0" maxOccurs="1" name="Location" type="coverageLocationType" />
          <xs:element minOccurs="0" maxOccurs="1" name="Bounds" type="coverageBoundsType" />
          <xs:element minOccurs="0" maxOccurs="1" name="Support" type="coverageSupportType" />
        </xs:sequence>
      </xs:extension>
    </xs:complexContent>
  </xs:complexType>

<xs:complexType name="spectralCharacterizationAxisType">
<xs:complexContent mixed="false">
 <xs:extension base="characterizationAxisType">
 <xs:sequence>
 <xs:element minOccurs="0" maxOccurs="1" name="ResPower" type="doubleParamType" />
 </xs:sequence>
 </xs:extension>
 </xs:complexContent>
</xs:complexType>

 <xs:complexType name="coverageLocationType">
  <xs:complexContent mixed="false">
     <xs:extension base="sedBaseCoordType">
      <xs:sequence>
       <xs:element minOccurs="0" maxOccurs="2" name="Value" type="doubleParamType" />
       <xs:element minOccurs="0" maxOccurs="1" name="Accuracy" type="accuracyType" />
       <xs:element minOccurs="0" maxOccurs="1" name="Resolution" type="doubleParamType" />
      </xs:sequence>
      </xs:extension>
   </xs:complexContent>
 </xs:complexType>
```

```
  <xs:complexType name="coverageBoundsType">
    <xs:complexContent mixed="false">
      <xs:extension base="Group">
        <xs:sequence>
          <xs:element minOccurs="0" maxOccurs="1" name="Extent" type="doubleParamType" />
           <xs:element minOccurs="0" maxOccurs="1" name="Range" type="intervalType" />
         </xs:sequence>
      </xs:extension>
    </xs:complexContent>
  </xs:complexType>


   <xs:complexType name="coverageSupportType">
    <xs:complexContent mixed="false">
      <xs:extension base="Group">
        <xs:sequence>
          <xs:element minOccurs="0" maxOccurs="1" name="Area" type="skyRegionType" />
          <xs:element minOccurs="0" maxOccurs="1" name="Extent" type="doubleParamType" />
          <xs:element minOccurs="0" maxOccurs="unbounded" name="Range" type="intervalType" />

        </xs:sequence>
      </xs:extension>
    </xs:complexContent>
  </xs:complexType>




<xs:complexType name="intervalType">
<xs:complexContent mixed="false">
<xs:extension base="Group">
<xs:sequence>
<xs:element minOccurs="0" maxOccurs="1" name="Min" type="doubleParamType" />
<xs:element minOccurs="0" maxOccurs="1" name="Max" type="doubleParamType" />
</xs:sequence>
</xs:extension>
</xs:complexContent>
</xs:complexType>

<xs:complexType name="samplingPrecisionType">
<xs:complexContent mixed="false">
<xs:extension base="Group">
<xs:sequence>
<xs:element minOccurs="0" maxOccurs="1" name="SamplingPrecisionRefVal" type="samplingPrecisionRefValType" />
<xs:element minOccurs="0" maxOccurs="1" name="SampleExtent" type="doubleParamType" />
</xs:sequence>
</xs:extension>
</xs:complexContent>
</xs:complexType>


<xs:complexType name="samplingPrecisionRefValType">
<xs:complexContent mixed="false">
<xs:extension base="Group">
<xs:sequence>
     <xs:element minOccurs="0" maxOccurs="1" name="FillFactor" type="doubleParamType" />
</xs:sequence>
</xs:extension>
</xs:complexContent>
</xs:complexType>
```

```
<xs:complexType name="skyRegionType">
<xs:complexContent mixed="false">
<xs:extension base="textParamType"/>
</xs:complexContent>
</xs:complexType>

<xs:complexType name="derivedDataType">
<xs:complexContent mixed="false">
<xs:extension base="Group">
<xs:sequence>
<xs:element minOccurs="0" maxOccurs="1" name="SNR" type="doubleParamType" />
<xs:element minOccurs="0" maxOccurs="1" name="VarAmpl" type="doubleParamType" />
<xs:element minOccurs="0" maxOccurs="1" name="Redshift" type="sedQuantityType" />
</xs:sequence>
</xs:extension>
</xs:complexContent>
</xs:complexType>


<xs:complexType name="dataIDType">
<xs:complexContent mixed="false">
<xs:extension base="Group">
<xs:sequence>
<xs:element minOccurs="0" maxOccurs="1" name="Title" type="textParamType" />
<xs:element minOccurs="0" maxOccurs="1" name="Creator" type="textParamType" />
<xs:element minOccurs="0" maxOccurs="unbounded" name="Collection" type="textParamType" />
<xs:element minOccurs="0" maxOccurs="1" name="DatasetID" type="textParamType" />
<xs:element minOccurs="0" maxOccurs="1" name="Date" type="dateParamType" />
<xs:element minOccurs="0" maxOccurs="1" name="Version" type="textParamType" />
<xs:element minOccurs="0" maxOccurs="1" name="Instrument" type="textParamType" />
<xs:element minOccurs="0" maxOccurs="1" name="CreationType" type="textParamType" />
<xs:element minOccurs="0" maxOccurs="1" name="Bandpass" type="textParamType" />
<xs:element minOccurs="0" maxOccurs="1" name="CreatorDID" type="textParamType" />
<xs:element minOccurs="0" maxOccurs="unbounded" name="Contributor" type="textParamType" />
<xs:element minOccurs="0" maxOccurs="1" name="Logo" type="textParamType" />
<xs:element minOccurs="0" maxOccurs="1" name="DataSource" type="textParamType" />
</xs:sequence>
</xs:extension>
</xs:complexContent>
</xs:complexType>
```

```
<xs:complexType name="targetType">
<xs:complexContent mixed="false">
<xs:extension base="Group">
<xs:sequence>
<xs:element minOccurs="0" maxOccurs="1" name="Name" type="textParamType" />
<xs:element minOccurs="0" maxOccurs="1" name="Description" type="textParamType" />
<xs:element minOccurs="0" maxOccurs="1" name="TargetClass" type="textParamType" />
<xs:element minOccurs="0" maxOccurs="1" name="SpectralClass" type="textParamType" />
<xs:element minOccurs="0" maxOccurs="1" name="Redshift" type="doubleParamType" />
<xs:element minOccurs="0" maxOccurs="1" name="Pos" type="positionParamType" />
<xs:element minOccurs="0" maxOccurs="1" name="VarAmpl" type="doubleParamType" />
<xs:element minOccurs="0" maxOccurs="1" name="CustomParams" type="arrayOfParamType" />
</xs:sequence>
</xs:extension>
</xs:complexContent>
</xs:complexType>




<xs:complexType name="arrayOfParamType">
<xs:sequence>
<xs:element minOccurs="0" maxOccurs="unbounded" name="Param" nillable="true" type="paramType" />
</xs:sequence>
</xs:complexType>


  <xs:attributeGroup name="STCReference">
    <xs:annotation>
      <xs:documentation>These four attributes represent the standard IVOA referencing system: internal (within th
    </xs:annotation>
    <xs:attribute name="id" type="xs:ID" use="optional"/>
    <xs:attribute name="idref" type="xs:IDREF" use="optional"/>
    <xs:attribute name="ucd" type="xs:string" use="optional"/>
    <xs:attribute ref="xlink:type" use="optional" default="simple"/>
    <xs:attribute ref="xlink:href" use="optional"/>
  </xs:attributeGroup>

  <xs:complexType name="coordSysType">
    <!--<xs:complexContent> -->
        <xs:sequence maxOccurs="unbounded">
          <xs:element ref="CoordFrame" minOccurs="0" maxOccurs="unbounded"/>
        </xs:sequence>
       <xs:attributeGroup ref="STCReference"/>
    <!-- </xs:complexContent>-->
  </xs:complexType>

  <xs:element name="CoordFrame" type="coordFrameType" abstract="true"/>
  <xs:element name="SpaceFrame" type="spaceFrameType" substitutionGroup="CoordFrame"/>
  <xs:element name="RedshiftFrame" type="redshiftFrameType" substitutionGroup="CoordFrame"/>
  <xs:element name="SpectralFrame" type="spectralFrameType" substitutionGroup="CoordFrame"/>
  <xs:element name="GenericCoordFrame" type="coordFrameType" substitutionGroup="CoordFrame"/>
  <xs:element name="TimeFrame" type="timeFrameType" substitutionGroup="CoordFrame"/>
```

```
  <xs:complexType name="coordFrameType">
  <xs:annotation>
    <xs:documentation>Simplification of STC version: RefPos is string</xs:documentation>
  </xs:annotation>
  <xs:sequence>
    <xs:element name="Name" type="xs:string" minOccurs="0"/>
    <xs:element name="ReferencePosition" type="xs:string" minOccurs="0"/>
  </xs:sequence>
  <xs:attribute name="id" type="xs:ID"/>
  <xs:attribute name="ucd" type="xs:string" use="optional"/>
</xs:complexType>

<xs:complexType name="spectralFrameType">
  <xs:complexContent>
    <xs:extension base="coordFrameType">
     <xs:sequence>
      <xs:element minOccurs="0" maxOccurs="1" name="Redshift" type="doubleParamType" />
     </xs:sequence>
    </xs:extension>
  </xs:complexContent>
</xs:complexType>



<xs:complexType name="timeFrameType">
  <xs:complexContent>
    <xs:extension base="coordFrameType">
     <xs:sequence>
      <xs:element minOccurs="0" maxOccurs="1" name="Zero" type="doubleParamType" />
     </xs:sequence>
    </xs:extension>
  </xs:complexContent>
</xs:complexType>


<xs:complexType name="redshiftFrameType">
  <xs:complexContent>
    <xs:extension base="coordFrameType">
      <xs:sequence>
        <xs:element name="DopplerDefinition" type="xs:string" nillable="true"/>
      </xs:sequence>
    </xs:extension>
  </xs:complexContent>
</xs:complexType>

<xs:complexType name="spaceFrameType">
  <xs:complexContent>
    <xs:extension base="coordFrameType">
      <xs:sequence>
       <xs:element minOccurs="0" maxOccurs="1" name="Equinox" type="doubleParamType" />
      </xs:sequence>
    </xs:extension>
  </xs:complexContent>
</xs:complexType>

</xs:schema>
```

## 7.2 Instance example

```
<?xml version="1.0" encoding="UTF-8" standalone="yes"?>
<Spectrum xmlns="http://www.ivoa.net/xml/Spectrum/Spectrum-1.01.xsd"
      xmlns:xsi="http://www.w3.org/2001/XMLSchema-instance"
      xsi:schemaLocation="http://www.ivoa.net/xml/Spectrum/Spectrum-1.01.xsd">
 <Char>
   <SpatialAxis name="Sky" ucd="pos.eq" unit="deg">
     <CoordSystem idref="ID000001"/>
     <Coverage>
      <Location><Value ucd="pos.eq.ra" unit="deg">132.4210</Value>
                <Value ucd="pos.eq.dec">12.1232</Value> </Location>
      <Bounds>
       <Extent ucd="pos.angDistance;instr.fov" unit="arcsec">20   </Extent>
      </Bounds>
     </Coverage>
     <Calibration>CALIBRATED</Calibration>
   </SpatialAxis>

   <TimeAxis name="Time" ucd="time" unit="d">
      <CoordSystem idref="ID000001"/>

     <Coverage>
     <Location><Value ucd="time.obs" unit="d">52148.3252</Value></Location>
     <Bounds>
      <Extent ucd="time.duration;obs.exposure" unit="s">1500.0</Extent>
     </Bounds>
    </Coverage>
    <Calibration>CALIBRATED</Calibration>
  </TimeAxis>

  <SpectralAxis name="SpectralCoord" ucd="em.wl" unit="Angstrom">
     <CoordSystem idref="ID000001"/>
    <Coverage>
     <Bounds>
      <Extent ucd="instr.bandwidth" unit="Angstrom">3000.0</Extent>
     </Bounds>
    </Coverage>
    <Accuracy>
      <BinLow ucd="em.wl;stat.min" unit="Angstrom"/>
      <BinHigh ucd="em.wl;stat.max" unit="Angstrom"/>
    </Accuracy>
    <Calibration>CALIBRATED</Calibration>
  </SpectralAxis>
   <!-- Flux density per unit wavelength  -->
   <FluxAxis name="Flux density" ucd="phot.flux.density;em.wl" unit="erg cm**(-2) s**(-1) Angstrom**(-1)">
     <Accuracy>
       <StatErrLow  unit="erg cm**(-2) s**(-1) Angstrom**(-1)"/>
       <StatErrHigh unit="erg cm**(-2) s**(-1) Angstrom**(-1)"/>
       <SysError>0.05</SysError>
     </Accuracy>
   </FluxAxis>

</Char>
```

```
<!-- STC metadata, with references to defined fields -->
 <CoordSys id="ID000001">
        <TimeFrame>
                <Name>UTC</Name>
<ReferencePosition>BARYCENTER</ReferencePosition>
        </TimeFrame>
        <SpaceFrame>
                <Name>ICRS</Name>
<ReferencePosition>BARYCENTER</ReferencePosition>
        </SpaceFrame>
        <SpectralFrame ucd="em.wl">
                <Name>Wavelength</Name>
<ReferencePosition>HELIOCENTER</ReferencePosition>
        </SpectralFrame>
        <GenericCoordFrame id="FD" ucd="phot.flux.density;em.wl">
                <Name>Flux density</Name>
        </GenericCoordFrame>
 </CoordSys>

  <Curation>
   <Publisher ucd="meta.curation">SAO</Publisher>
   <PublisherID ucd="meta.ref.url;meta.curation">ivo://cfa.harvard.edu</PublisherID>
   <Contact>
     <Name ucd="meta.bib.author;meta.curation">Jonathan McDowell</Name>
     <Email ucd="meta.email">jcm@cfa.harvard.edu</Email>
   </Contact>
  </Curation>

  <DataID>
    <Title>Arp 220 SED</Title>
    <Creator ucd="meta.curation.creator">SAO/FLWO</Creator>
    <DatasetID>ivo://cfa.harvard.edu/flwo#10314</DatasetID>
    <Collection>Archival</Collection>
    <Date ucd="time.epoch;meta.dataset">2003-12-31T14:00:02Z</Date>
    <Version ucd="meta.version;meta.dataset">1</Version>
    <Instrument ucd="meta.id;instr">BCS</Instrument>
    <Logo ucd="meta.ref.url">http://cfa-www.harvard.edu/nvo/cfalogo.jpg</Logo>
  </DataID>

  <Derived>
    <SNR>3.0</SNR>
  </Derived>
```

```
  <!-- Use table structure -->

  <ArrayOfPoint>
   <Point>
    <SpectralAxis>
     <Value>3200.0</Value>
     <Accuracy><BinLow>3195.0</BinLow><BinHigh>3205.0</BinHigh></Accuracy>
    </SpectralAxis>
    <FluxAxis>
      <Value>1.38E-12</Value>
      <Accuracy><StatErrLow>5.2E-14</StatErrLow><StatErrHigh>6.2E-14</StatErrHigh></Accuracy>
      <Quality>0</Quality>
    </FluxAxis>
   </Point>

   <Point>
    <SpectralAxis>
     <Value>3210.5</Value>
     <Accuracy><BinLow>3205.0</BinLow><BinHigh>3216.0</BinHigh></Accuracy>
    </SpectralAxis>
    <FluxAxis>
      <Value>1.12E-12</Value>
      <Accuracy><StatErrLow>1.12E-12</StatErrLow><StatErrHigh>0</StatErrHigh></Accuracy>
      <Quality>0</Quality>
    </FluxAxis>
   </Point>

   <Point>
    <SpectralAxis>
     <Value>3222.0</Value>
     <Accuracy><BinLow>3216.0</BinLow><BinHigh>3228.0</BinHigh></Accuracy>
    </SpectralAxis>
    <FluxAxis>
      <Value>1.42E-12</Value>
      <Accuracy><StatErrLow>1.3E-14</StatErrLow><StatErrHigh>0.2E-14</StatErrHigh></Accuracy>
      <Quality>3</Quality>
    </FluxAxis>
   </Point>

  </ArrayOfPoint>

  <!-- OR -->

  <ArrayOfFlatPoint>
  <Point SP="3200.0" SP_BinL="3195.0" SP_BinH="3205.0" F="1.48E-12" F_ErrL="2.0E-14"  F_ErrH="2.0E-14"/>
  <Point SP="3210.0" SP_BinL="3205.0" SP_BinH="3215.0" F="1.48E-12" F_ErrL="3.2E-14"  F_ErrH="3.8E-14"/>
  <Point SP="3220.0" SP_BinL="3215.0" SP_BinH="3225.0" F="1.48E-12" F_ErrL="1.48E-12"  F_ErrH="0.0" />

 </ArrayOfFlatPoint>
</Spectrum>
```

# Part 3: VOTABLE serialization

# 8 VOTABLE serialization

## 8.1 Mapping Schema to VOTABLE

We reproduce below the XML schema instance example as a VOTABLE instance example. To go from the XML instance to the VOTABLE instance, we:

- - map the top level element to a RESOURCE
- - map all elements with simple content to PARAM
- - map all elements with complex content to GROUP
- - map the element names (with appropriate path) to values of the utype attribute,
- - but, handle the FIELDS and Data elements in a special way. The FIELDS element is used to define the table fields and the Data element is used to define the table data.
- - but, also, all the second level elements below RESOURCE except SPECTRUM map to an initial TABLE, while we map SPECTRUM to a second TABLE.
- - most of the elements extend the Param element, to which I have added an optional name attribute that I have not used in the instance. If this attribute is used, it can hold the name attributes of the PARAM and FIELD; otherwise the relevant attributes could be filled with the same value as the utype (without namespace prefix).

How can this be generalized to mapping an arbitrary data model schema to VOTABLE? The only tricky parts are

- **Spotting where the tabledata parts are.** We could require any DM schema that maps to VOTABLE to include elements called FIELDS and Data (perhaps ROWS would be a better name), otherwise you would get a VOTABLE with no data section.
- **Spotting where to start the main TABLE (i.e. the fact that SPECTRUM is special).** We could change the schema to have an explicit attribute, annotation or other marker to tell us this.

These issues will require further discussion for future models.

## 8.2 A VOTABLE instance

The VOTable version of Spectrum uses a single VOTable <TABLE> (Note that this may appear as one of many tables within an SED VOTable). The data model fields described above as arrays map to VOTable FIELDs, while the remaining fields map to PARAM.

We use nested GROUP constructs to delimit data model objects within the main object, and PARAM and FIELD tags for attributes. The nesting beyond a single GROUP is optional, as for cases for which the utypes are unique within a group, the utypes can be used to infer the data-model structure. See http://webtest.aoc.nrao.edu/ivoa-dal for a service returning VOTABLE Spectrum instances with only one level of GROUP.

Names of fields and parameters are left to the data provider. The utype and ucd attributes are used to denote data model and UCD tags. The schema and namespace for the utypes is the XML schema given in section 8.4. We have made up arbitrary NAME attributes for the PARAM and these are not to be considered standard; the name fields are free to be whatever the data provider wants, allowing compatibility with local archive nomenclature. The NAME attributes for the FIELD elements are also not standardized (of course they must be the same as in the matching FIELDrefs); it is the utype attribute which is standardized.

The one departure from the XML schema below is that the 'Data' element and the individual 'Point' elements are implicitly represented by the table structure itself. Perhaps a UTYPE attribute to the TABLEDATA element could be used to make this explicit.

The examples below describe a single SPECTRUM.

```
<?xml version="1.0" encoding="UTF-8"?>
<VOTABLE version="1.1"
  xmlns:xsi="http://www.w3.org/2001/XMLSchema-instance"
  xsi:noNamespaceSchemaLocation="xmlns:http://www.ivoa.net/xml/VOTable/VOTable-1.1.xsd"
  xmlns:spec="http://www.ivoa.net/xml/SpectrumModel/v1.01"
  xmlns="http://www.ivoa.net/xml/VOTable/v1.1">
<RESOURCE utype="spec:Spectrum">

<TABLE utype="spec:Spectrum">
<GROUP utype="spec:Target">
 <PARAM name="Target" utype="spec:Target.Name" datatype="char" arraysize="*" value="Arp 220"/>
 <PARAM name="TargetPos" utype="spec:Target.pos" unit="deg" datatype="double"
                                arraysize="2" value="233.737917 23.503330"/>
 <PARAM name="z" utype="spec:Target.redshift" datatype="float" value="0.0018"/>
</GROUP>

<!-- SegmentType can be Photometry, TimeSeries or Spectrum -->
<PARAM name="Segtype" utype="spec:SegmentType" datatype="char" arraysize="*"
                                                  value="Photometry" ucd="meta.code"/>
<GROUP name="CoordSys" utype="spec:CoordSys">
 <GROUP utype="CoordSys.SpaceFrame">
   <PARAM name="System" utype="spec:CoordSys.SpaceFrame.Name" ucd="pos.frame"
                                        datatype="char" arraysize="*" value="ICRS"/>
   <PARAM name="Equinox" utype="spec:CoordSys.SpaceFrame.Equinox" ucd="time.equinox;pos.eq"
                                        datatype="float" value="2000.0" />
 </GROUP>
 <GROUP utype="spec:CoordSys.TimeFrame">
  <PARAM name="TimeFrame" utype="spec:CoordSys.TimeFrame.Name" ucd="time.scale" datatype="char"
                                      arraysize="*" value="UTC"/>
 </GROUP>
 <GROUP utype="spec:CoordSys.SpectralFrame">
  <PARAM name="SpectralFrame" utype="spec:CoordSys.SpectralFrame.RefPos" ucd="sdm:spect.frame"
                                     datatype="char" arraysize="*" value="BARYCENTER"/>
 </GROUP>
</GROUP>
```

```
<GROUP utype="spec:Char">
 <GROUP utype="spec:Char.SpatialAxis">
   <PARAM name="SpatialAxisName" utype="name" ucd="pos.eq" unit="deg" value="Sky"/>
   <GROUP utype="spec:Char.SpatialAxis.Coverage">
    <GROUP utype="spec:Char.SpatialAxis.Coverage.Location">
     <PARAM name="SkyPos" utype="Char.SpatialAxis.Coverage.Location.Value"
                                    ucd="pos.eq" unit="deg"
                                    datatype="double" arraysize="2" value="132.4210 12.1232"/>
   </GROUP>
   <GROUP utype="Bounds">
     <PARAM name="SkyExtent" utype="Char.SpatialAxis.Coverage.Extent" ucd="pos.angDistance;instr.fov"
                      datatype="double" unit="arcsec" value="20"/>
   </GROUP>
  </GROUP>
 </GROUP>



 <GROUP utype="spec:Char.TimeAxis">
  <PARAM name="TimeAxisName" utype="Char.TimeAxis.Name" ucd="time" unit="d" value="Time"/>
  <GROUP utype="Char.TimeAxis.Coverage">
   <GROUP utype="Char.TimeAxis.Coverage.Location">
    <PARAM name="TimeObs" utype="Char.TimeAxis.Coverage.Location.Value" ucd="time.epoch;obs"
                                    datatype="double" value="52148.3252"/>
   </GROUP>
   <GROUP utype="Char.TimeAxis.Coverage.Bounds">
    <PARAM name="TimeExtent" utype="Char.TimeAxis.Coverage.Bounds.Extent" ucd="time.duration"
                      unit="s" datatype="double" value="1500.0" />
    <PARAM name="TimeStart" utype="Char.TimeAxis.Coverage.Bounds.Start" ucd="time.start" unit="s"
                                      datatype="double" value="52100.000" />
    <PARAM name="TimeStop" utype="Char.TimeAxis.Coverage.Bounds.Stop" ucd="time.end" unit="s"
                                      datatype="double" value="52300.000" />
   </GROUP>
   <GROUP utype="Char.TimeAxis.Coverage.Support">
    <PARAM name="TimeExtent" utype="Char.TimeAxis.Coverage.Support.Extent" ucd="time.duration;obs.exposure"
                      unit="s" datatype="double" value="1500.0" />
    <PARAM name="TimeStart" utype="Char.TimeAxis.Coverage.Bounds.Start" ucd="time.start" unit="s"
                                      datatype="double" value="52100.000" />
    <PARAM name="TimeStop" utype="Char.TimeAxis.Coverage.Bounds.Stop" ucd="time.end" unit="s"
                                      datatype="double" value="52300.000" />
   </GROUP>

  </GROUP>
 </GROUP>

 <GROUP utype="spec:Char.SpectralAxis">
  <PARAM name="SpectralAxisName" utype="Char.SpectralAxis.Name" ucd="em.wl" unit="angstrom" value="Wavelength"/>
  <GROUP utype="Char.SpectralAxis.Coverage">
   <GROUP utype="Char.SpectralAxis.Coverage.Bounds">
    <PARAM name="SpectralExtent" utype="Char.SpectralAxis.Coverage.Bounds.Extent" ucd="instr.bandwidth"
                      unit="angstrom" datatype="double" value="3000.0"/>
   </GROUP>
  </GROUP>
 </GROUP>
</GROUP>
```

```
<GROUP utype="spec:Curation">
 <PARAM name="Publisher" utype="spec:Curation.Publisher" ucd="meta.curation"
                      datatype="char" arraysize="*" value="SAO"/>
 <PARAM name="PubID" utype="spec:Curation.PublisherID" ucd="meta.ref.url;meta.curation" datatype="char"
                                      arraysize="*" value="ivo://cfa.harvard.edu"/>
 <PARAM name="Contact" utype="spec:Curation.Contact.Name" ucd="meta.bib.author;meta.curation"
                      datatype="char" arraysize="*" value="Jonathan McDowell"/>
 <PARAM name="email" utype="spec:Curation.Contact.Email" ucd="meta.email" datatype="char"
                                     arraysize="*" value="jcm@cfa.harvard.edu"/>
</GROUP>


<GROUP utype="spec:DataID">
 <PARAM name="Title" utype="spec:DataID.Title" datatype="char" arraysize="*" value="Arp 220 SED"/>
 <PARAM name="Creator" utype="spec:Segment.DataID.Creator" ucd="meta.curation" datatype="char"
                                     arraysize="*" value="ivo://sao/FLWO"/>
 <PARAM name="DataDate" utype="spec:DataID.Date" ucd="time.epoch;meta.dataset"
                     datatype="char" arraysize="*" value="2003-12-31T14:00:02Z"/>
 <PARAM name="Version" utype="spec:DataID.Version" ucd="meta.version;meta.dataset"
                     datatype="char" arraysize="*" value="1"/>
 <PARAM name="Instrument" utype="spec:DataID.Instrument" ucd="meta.id;instr" datatype="char"
                                   arraysize="*" value="BCS"/>
 <PARAM name="Filter" utype="spec:DataID.Collection" ucd="inst.filter.id" datatype="char"
                                  arraysize="*" value="G300"/>
 <PARAM name="CreationType" utype="spec:DataID.CreationType" datatype="char" arraysize="*" value="Archival"/>
 <PARAM name="Logo" utype="spec:DataID.Logo" ucd="meta.ref.url" datatype="char"
                                       arraysize="*" value="http://cfa-www.harvard.edu/nvo/cfalogo.jpg"/>
</GROUP>

<GROUP utype="spec:Derived">
 <PARAM name="SNR" utype="spec:Derived.SNR" datatype="float" value="3.0"/>
</GROUP>

<GROUP utype="spec:Data">

<GROUP utype="spec:Data.SpectralAxis">
 <FIELDref ref="Coord"/>

 <GROUP utype="spec:Data.SpectralAxis.Accuracy">
  <FIELDref ref="BinLow"/>
  <FIELDref ref="BinHigh"/>
 </GROUP>
<!-- In this case Resolution is demoted from Field to Param since it is constant -->
 <PARAM name="Resolution" utype="spec:Data.SpectralAxis.Resolution"
      ucd="spect.resolution;em.wl"    unit="angstrom" datatype="float" value="14.2"/>
</GROUP>
```

```
<GROUP utype="spec:Data.FluxAxis">
 <FIELDref ref="Flux1"/>
 <GROUP utype="spec:Data.FluxAxis.Accuracy">
  <FIELDref ref="ErrorLow"/>
  <FIELDref ref="ErrorHigh"/>
  <PARAM name="SysErr" utype="SysErr" unit="" datatype="float" value="0.05"/>
 </GROUP>
 <FIELDref ref="Quality"/>
</GROUP>
</GROUP>

<FIELD name="Coord" ID="Coord" utype="spec:Data.SpectralAxis.Value" ucd="em.wl"
              datatype="double" unit="angstrom"/>
<FIELD name="BinLow" ID="BinLow" utype="spec:Data.SpectralAxis.BinLow"
              ucd="em.wl;stat.min"
              datatype="double" unit="angstrom"/>
<FIELD name="BinHigh" ID="BinHigh" utype="spec:Data.SpectralAxis.BinHigh"
               ucd="em.wl;stat.max"
              datatype="double" unit="angstrom"/>
<FIELD name="Flux" ID="Flux1" utype="spec:Data.FluxAxis.value" ucd="phot.flux.density;em.wl"
              datatype="double" unit="erg cm**(-2) s**(-1) angstrom**(-1)"/>
<FIELD name="ErrorLow" ID="ErrorLow" utype="spec:Data.FluxAxis.Accuracy.StatErrLow"
              datatype="double" unit="erg cm**(-2) s**(-1) angstrom**(-1)"/>
<FIELD name="ErrorHigh" ID="ErrorHigh" utype="spec:Data.FluxAxis.Accuracy.StatErrHigh"
              datatype="double" unit="erg cm**(-2) s**(-1) angstrom**(-1)"/>
<FIELD name="Quality" ID="Quality" datatype="int" utype="spec:Data.FluxAxis.Quality"/>
<DATA>
<TABLEDATA>
<!-- Note slightly nonlinear wavelength solution -->
<!-- Second row is upper limit -->
<!-- Third row has quality mask set -->
<TR><TD>3200.0</TD><TD>3195.0</TD><TD>3205.0</TD><TD>1.38E-12</TD><TD>5.2E-14</TD><TD>6.2E-14</TD>
                                                                                     <TD>0</TD></TR>
<TR><TD>3210.5</TD><TD>3205.0</TD><TD>3216.0</TD><TD>1.12E-12</TD><TD>1.12E-12</TD>
                                                                      <TD>0</TD><TD>0</TD></TR>
<TR><TD>3222.0</TD><TD>3216.0</TD><TD>3228.0</TD><TD>1.42E-12</TD><TD>1.3E-14</TD>
                                                                      <TD>0.2E-14</TD><TD>3</TD></TR>
</TABLEDATA>
</DATA>
</TABLE>
</RESOURCE>
</VOTABLE>
```

A second example, based on the reference SSAP proxy service for the JHU SDSS spectrum archive:

```xml
<?xml version="1.0" encoding="UTF-8"?>
<VOTABLE xmlns:xsi="http://www.w3.org/2001/XMLSchema-instance"
   xsi:noNamespaceSchemaLocation="xmlns:http://www.ivoa.net/xml/VOTable/VOTable-1.1.xsd"
   xmlns:spec="http:www.ivoa.net/xml/SpectrumModel/v1.0" version="1.1">
<RESOURCE utype="spec:Spectrum">
<DESCRIPTION>Spectrum dataset generated by DALServer</DESCRIPTION>
<TABLE utype="spec:Spectrum">
<FIELD ID="DataSpectralValue" name="DataSpectralValue" datatype="double" ucd="em.wl" utype="spec:Spectrum.Data.SpectralAxis.Value" unit="angstrom">
<DESCRIPTION>Spectral coordinates for points</DESCRIPTION>
</FIELD>
<FIELD ID="DataSpectralBinLow" name="DataSpectralBinLow" datatype="double" ucd="em.wl;stat.min" utype="spec:Spectrum.Data.SpectralAxis.Accuracy.BinLow" unit="angstrom">
<DESCRIPTION>Spectral coord bin lower end</DESCRIPTION>
</FIELD>
<FIELD ID="DataSpectralBinHigh" name="DataSpectralBinHigh" datatype="double" ucd="em.wl;stat.max" utype="spec:Spectrum.Data.SpectralAxis.Accuracy.BinHigh" unit="angstrom">
<DESCRIPTION>Spectral coord bin upper end</DESCRIPTION>
</FIELD>
<FIELD ID="DataFluxValue" name="DataFluxValue" datatype="double" ucd="phot.flux.density;em.wl" utype="spec:Spectrum.Data.FluxAxis.Value"
   unit="10**(-17) erg/cm**2/s/angstrom">
<DESCRIPTION>Flux values for points</DESCRIPTION>
</FIELD>
<FIELD ID="DataFluxStatErrLow" name="DataFluxStatErrLow" datatype="double" ucd="phot.flux.density;em.wl" utype="spec:Spectrum.Data.FluxAxis.Accuracy.StatErrLow"
   unit="10**(-17) erg/cm**2/s/angstrom"><DESCRIPTION>Flux lower error</DESCRIPTION>
</FIELD>
<FIELD ID="DataFluxStatErrHigh" name="DataFluxStatErrHigh" datatype="double" ucd="phot.flux.density;em.wl" utype="spec:Spectrum.Data.FluxAxis.Accuracy.StatErrHigh"
 unit="10**(-17) erg/cm**2/s/angstrom"> <DESCRIPTION>Flux upper error</DESCRIPTION>
</FIELD>
<FIELD ID="DataFluxQuality" name="DataFluxQuality" datatype="long" ucd="meta.code.qual;phot.flux" utype="spec:Spectrum.Data.FluxAxis.Quality">
<DESCRIPTION>Flux measurement quality mask</DESCRIPTION>
</FIELD>
<GROUP ID="Spectrum" name="Spectrum" utype="spec:Spectrum">
<DESCRIPTION>General Dataset Metadata</DESCRIPTION>
<PARAM ID="DataModel" datatype="char" name="DataModel" utype="spec:Spectrum.DataModel" value="Spectrum 1.0" arraysize="*">
<DESCRIPTION>Datamodel name and version</DESCRIPTION>
</PARAM>
<PARAM ID="DatasetType" datatype="char" name="DatasetType" utype="spec:Spectrum.Type" value="Spectrum" arraysize="*">
<DESCRIPTION>Dataset or segment type</DESCRIPTION>
</PARAM>
<PARAM ID="DataLength" datatype="long" name="DataLength" utype="spec:Spectrum.Length" value="3820">
<DESCRIPTION>Number of points</DESCRIPTION>
</PARAM>
</GROUP>
<GROUP ID="DataID" name="DataID" utype="spec:DataID">
<DESCRIPTION>Dataset Identification Metadata</DESCRIPTION>
<PARAM ID="Title" datatype="char" name="Title" ucd="meta.title;meta.dataset" utype="spec:Spectrum.DataID.Title"
 value="SDSS J142906.30+016000.00 Rosat_d 0535-51998-01" arraysize="*"><DESCRIPTION>Dataset Title</DESCRIPTION>
</PARAM>
<PARAM ID="Creator" datatype="char" name="Creator" utype="spec:Spectrum.DataID.Creator" value="sdss" arraysize="*">
<DESCRIPTION>Dataset creator</DESCRIPTION>
</PARAM>
<PARAM ID="Collection" datatype="char" name="Collection" utype="spec:Spectrum.DataID.Collection" value="ivo://sdss/dr5/spec" arraysize="*">
<DESCRIPTION>Data collection to which dataset belongs</DESCRIPTION>
</PARAM>
```

```
<PARAM ID="CreatorDID" datatype="char" name="CreatorDID" ucd="meta.id" utype="spec:Spectrum.DataID.CreatorDID" value="ivo://sdss/dr5/spec#150812447593201664"
 arraysize="*"><DESCRIPTION>Creator's ID for the dataset</DESCRIPTION>
</PARAM>
<PARAM ID="CreatorDate" datatype="char" name="CreatorDate" ucd="time;meta.dataset" utype="spec:Spectrum.DataID.Date" value="2001-03-31T09:10:18.8600000-04:00"
 arraysize="*"><DESCRIPTION>Data processing/creation date</DESCRIPTION>
</PARAM>
<PARAM ID="CreatorVersion" datatype="char" name="CreatorVersion" ucd="meta.version;meta.dataset" utype="spec:Spectrum.DataID.Version" value="3.36.10"
 arraysize="*"><DESCRIPTION>Version of dataset</DESCRIPTION>
</PARAM>
<PARAM ID="Instrument" datatype="char" name="Instrument" ucd="meta.id;instr" utype="spec:Spectrum.DataID.Instrument" value="SDSS 2.5-M SPEC1 v4_7" arraysize="*">
<DESCRIPTION>Instrument name</DESCRIPTION>
</PARAM>
<PARAM ID="DataSource" datatype="char" name="DataSource" utype="spec:Spectrum.DataID.DataSource" value="survey" arraysize="*">
<DESCRIPTION>Original source of the data</DESCRIPTION>
</PARAM>
<PARAM ID="CreationType" datatype="char" name="CreationType" utype="spec:Spectrum.DataID.CreationType" value="Archival" arraysize="*">
<DESCRIPTION>Dataset creation type</DESCRIPTION>
</PARAM>
</GROUP>
<GROUP ID="Curation" name="Curation" utype="spec:Curation">
<DESCRIPTION>Curation Metadata</DESCRIPTION>
<PARAM ID="Publisher" datatype="char" name="Publisher" ucd="meta.curation" utype="spec:Spectrum.Curation.Publisher" value="DALServer Proxy" arraysize="*">
<DESCRIPTION>Dataset publisher</DESCRIPTION>
</PARAM>
<PARAM ID="PublisherDID" datatype="char" name="PublisherDID" ucd="meta.ref.url;meta.curation" utype="spec:Spectrum.Curation.PublisherDID"
 value="ivo://jhu/sdss/dr5#150812447593201664" arraysize="*"><DESCRIPTION>Publisher's ID for the dataset ID</DESCRIPTION>
</PARAM>
<PARAM ID="Rights" datatype="char" name="Rights" utype="spec:Spectrum.Curation.Rights" value="public" arraysize="*">
<DESCRIPTION>Restrictions on data access</DESCRIPTION>
</PARAM>
</GROUP>
<GROUP ID="Target" name="Target" utype="spec:Target">
<DESCRIPTION>Target Metadata</DESCRIPTION>
<PARAM ID="TargetName" datatype="char" name="TargetName" ucd="meta.id;src" utype="spec:Spectrum.Target.Name" value="SDSS J142906.30+016000.00" arraysize="*">
<DESCRIPTION>Target name</DESCRIPTION>
</PARAM>
<PARAM ID="TargetDescription" datatype="char" name="TargetDescription" utype="spec:Spectrum.Target.Description" value="0535-51998-01" arraysize="*">
<DESCRIPTION>Target description</DESCRIPTION>
</PARAM>
<PARAM ID="TargetClass" datatype="char" name="TargetClass" ucd="src.class" utype="spec:Spectrum.Target.Class" value="Rosat_d" arraysize="*">
<DESCRIPTION>Object class of observed target</DESCRIPTION>
</PARAM>
<PARAM ID="TargetPos" unit="deg" datatype="double" name="TargetPos" ucd="pos.eq;src" utype="spec:Spectrum.Target.Pos" value="217.276250 1.289608" arraysize="2">
<DESCRIPTION>Target RA and Dec</DESCRIPTION>
</PARAM>
<PARAM ID="SpectralClass" datatype="char" name="SpectralClass" ucd="src.spType" utype="spec:Spectrum.Target.SpectralClass" value="Star" arraysize="*">
<DESCRIPTION>Object spectral class</DESCRIPTION>
</PARAM>
<PARAM ID="Redshift" datatype="double" name="Redshift" ucd="src.redshift" utype="spec:Spectrum.Target.Redshift" value="0.000197439">
<DESCRIPTION>Target redshift</DESCRIPTION>
</PARAM>
<PARAM ID="VarAmpl" datatype="float" name="VarAmpl" ucd="src.var.amplitude" utype="spec:Spectrum.Target.VarAmpl" value="0">
<DESCRIPTION>Target variability amplitude (typical)</DESCRIPTION>
</PARAM>
</GROUP>
```

```
<GROUP ID="Derived" name="Derived" utype="spec:Derived">
<DESCRIPTION>Derived Metadata</DESCRIPTION>
<PARAM ID="DerivedRedshift" datatype="double" name="DerivedRedshift" utype="spec:Spectrum.Derived.Redshift.Value" value="0.000197439">
<DESCRIPTION>Measured redshift for spectrum</DESCRIPTION>
</PARAM>
<PARAM ID="RedshiftStatError" datatype="float" name="RedshiftStatError" ucd="stat.error;src.redshift" utype="spec:Spectrum.Derived.Redshift.StatError"
   value="0.000166012"><DESCRIPTION>Error on measured redshift</DESCRIPTION>
</PARAM>
<PARAM ID="RedshiftConfidence" datatype="float" name="RedshiftConfidence" utype="spec:Spectrum.Derived.Redshift.Confidence" value="0.360793">
<DESCRIPTION>Confidence value on redshift</DESCRIPTION>
</PARAM>
<PARAM ID="DerivedVarAmpl" datatype="float" name="DerivedVarAmpl" ucd="src.var.amplitude;arith.ratio" utype="spec:Spectrum.Derived.VarAmpl" value="0">
<DESCRIPTION>Variability amplitude as fraction of mean</DESCRIPTION>
</PARAM>
</GROUP>
<GROUP ID="CoordSys" name="CoordSys" utype="spec:CoordSys">
<DESCRIPTION>Coordinate System Metadata</DESCRIPTION>
<PARAM ID="SpaceFrameName" datatype="char" name="SpaceFrameName" utype="spec:Spectrum.CoordSys.SpaceFrame.Name" value="FK5" arraysize="*">
<DESCRIPTION>Spatial coordinate frame name</DESCRIPTION>
</PARAM>
<PARAM ID="SpaceFrameUcd" datatype="char" name="SpaceFrameUcd" utype="spec:Spectrum.CoordSys.SpaceFrame.Ucd" value="pos.eq" arraysize="*">
<DESCRIPTION>Space frame UCD</DESCRIPTION>
</PARAM>
<PARAM ID="SpaceFrameEquinox" unit="yr" datatype="double" name="SpaceFrameEquinox" ucd="time.equinox;pos.frame"
 utype="spec:Spectrum.CoordSys.SpaceFrame.Equinox" value="2000"><DESCRIPTION>Equinox</DESCRIPTION>
</PARAM>
<PARAM ID="TimeFrameName" datatype="char" name="TimeFrameName" ucd="time.scale" utype="spec:Spectrum.CoordSys.TimeFrame.Name" value="TAI" arraysize="*">
<DESCRIPTION>Timescale</DESCRIPTION>
</PARAM>
<PARAM ID="TimeFrameZero" unit="d" datatype="double" name="TimeFrameZero" ucd="time;arith.zp" utype="spec:Spectrum.CoordSys.TimeFrame.Zero" value="0">
<DESCRIPTION>Zero point of timescale in MJD</DESCRIPTION>
</PARAM>
<PARAM ID="TimeFrameRefPos" datatype="char" name="TimeFrameRefPos" ucd="sdm:time.frame" utype="spec:Spectrum.CoordSys.TimeFrame.RefPos" value="Topocentric"
 arraysize="*"><DESCRIPTION>Location for times of photon arrival</DESCRIPTION>
</PARAM>
<PARAM ID="SpectralFrameUcd" datatype="char" name="SpectralFrameUcd" utype="spec:Spectrum.CoordSys.SpectralFrame.Ucd" value="em.wl" arraysize="*">
<DESCRIPTION>Spectral frame UCD</DESCRIPTION>
</PARAM>
<PARAM ID="SpectralFrameRefPos" datatype="char" name="SpectralFrameRefPos" utype="spec:Spectrum.CoordSys.SpectralFrame.RefPos" value="Topocentric" arraysize="*">
<DESCRIPTION>Spectral frame origin</DESCRIPTION>
</PARAM>
</GROUP>
```

```
<GROUP ID="Char.SpatialAxis" name="Char.SpatialAxis" utype="spec:Char.SpatialAxis">
<DESCRIPTION>Spatial Axis Characterization</DESCRIPTION>
<PARAM ID="SpatialAxisName" datatype="char" name="SpatialAxisName" utype="spec:Spectrum.Char.SpatialAxis.Name" value="Sky" arraysize="*">
<DESCRIPTION>Name for spatial axis</DESCRIPTION>
</PARAM>
<PARAM ID="SpatialAxisUcd" datatype="char" name="SpatialAxisUcd" utype="spec:Spectrum.Char.SpatialAxis.Ucd" value="pos.eq" arraysize="*">
<DESCRIPTION>UCD for spatial coord</DESCRIPTION>
</PARAM>
<PARAM ID="SpatialAxisUnit" datatype="char" name="SpatialAxisUnit" utype="spec:Spectrum.Char.SpatialAxis.Unit" value="deg" arraysize="*">
<DESCRIPTION>Unit for spatial coord</DESCRIPTION>
</PARAM>
<PARAM ID="SpatialLocation" unit="deg" datatype="double" name="SpatialLocation" ucd="pos.eq" utype="spec:Spectrum.Char.SpatialAxis.Coverage.Location.Value"
  value="217.276250 1.289608" arraysize="2"><DESCRIPTION>Spatial Position</DESCRIPTION>
</PARAM>
<PARAM ID="SpatialExtent" unit="deg" datatype="double" name="SpatialExtent" ucd="instr.fov" utype="spec:Spectrum.Char.SpatialAxis.Coverage.Bounds.Extent"
  value="0.00083333333333333339"><DESCRIPTION>Aperture angular size</DESCRIPTION>
</PARAM>
<PARAM ID="SpatialCalibration" datatype="char" name="SpatialCalibration" ucd="meta.code.qual" utype="spec:Spectrum.Char.SpatialAxis.Accuracy.Calibration"
 value="calibrated" arraysize="*"><DESCRIPTION>Type of spatial coord calibration</DESCRIPTION>
</PARAM>
</GROUP>
<GROUP ID="Char.SpectralAxis" name="Char.SpectralAxis" utype="spec:Char.SpectralAxis">
<DESCRIPTION>Spectral Axis Characterization</DESCRIPTION>
<PARAM ID="SpectralAxisName" datatype="char" name="SpectralAxisName" utype="spec:Spectrum.Char.SpectralAxis.Name" value="SpectralCoord" arraysize="*">
<DESCRIPTION>Name for spectral axis</DESCRIPTION>
</PARAM>
<PARAM ID="SpectralAxisUcd" datatype="char" name="SpectralAxisUcd" utype="spec:Spectrum.Char.SpectralAxis.Ucd" value="em.wl" arraysize="*">
<DESCRIPTION>UCD for spectral coord</DESCRIPTION>
</PARAM>
<PARAM ID="SpectralAxisUnit" datatype="char" name="SpectralAxisUnit" utype="spec:Spectrum.Char.SpectralAxis.Unit" value="angstrom" arraysize="*">
<DESCRIPTION>Unit for spectral coord</DESCRIPTION>
</PARAM>
<PARAM ID="SpectralLocation" datatype="double" name="SpectralLocation" ucd="instr.bandpass" utype="spec:Spectrum.Char.SpectralAxis.Coverage.Location.Value"
  value="6515.408759029946"><DESCRIPTION>Spectral coord value</DESCRIPTION>
</PARAM>
<PARAM ID="SpectralExtent" datatype="double" name="SpectralExtent" ucd="instr.bandwidth" utype="spec:Spectrum.Char.SpectralAxis.Coverage.Bounds.Extent"
 value="5386.6535232062424"><DESCRIPTION>Width of spectrum</DESCRIPTION>
</PARAM>
<PARAM ID="SpectralStart" datatype="double" name="SpectralStart" ucd="em.wl;stat.min" utype="spec:Spectrum.Char.SpectralAxis.Coverage.Bounds.Start"
 value="3822.0819974268243"><DESCRIPTION>Start in spectral coordinate</DESCRIPTION>
</PARAM>
<PARAM ID="SpectralStop" datatype="double" name="SpectralStop" ucd="em.wl;stat.max" utype="spec:Spectrum.Char.SpectralAxis.Coverage.Bounds.Stop"
 value="9208.7355206330667"><DESCRIPTION>Stop in spectral coordinate</DESCRIPTION>
</PARAM>
<PARAM ID="SpectralBinSize" datatype="double" name="SpectralBinSize" ucd="spect.binSize" utype="spec:Spectrum.Char.SpectralAxis.Accuracy.BinSize" value="0">
<DESCRIPTION>Spectral coord bin size</DESCRIPTION>
</PARAM>
<PARAM ID="SpectralStatError" datatype="double" name="SpectralStatError" ucd="stat.error;em.wl" utype="spec:Spectrum.Char.SpectralAxis.Accuracy.StatError" value="0">
<DESCRIPTION>Spectral coord statistical error</DESCRIPTION>
</PARAM>
```

```
<PARAM ID="SpectralSysError" datatype="double" name="SpectralSysError" ucd="stat.error;em.wl" utype="spec:Spectrum.Char.SpectralAxis.Accuracy.SysError" value="0">
<DESCRIPTION>Spectral coord systematic error</DESCRIPTION>
</PARAM>
<PARAM ID="SpectralCalibration" datatype="char" name="SpectralCalibration" ucd="meta.code.qual" utype="spec:Spectrum.Char.SpectralAxis.Accuracy.Calibration"
  value="Absolute" arraysize="*"><DESCRIPTION>Type of spectral coord calibration</DESCRIPTION>
</PARAM>
<PARAM ID="SpectralResolution" datatype="double" name="SpectralResolution" ucd="spect.resolution" utype="spec:Spectrum.Char.SpectralAxis.Resolution" value="0">
<DESCRIPTION>Spectral resolution FWHM</DESCRIPTION>
</PARAM>
</GROUP>
<GROUP ID="Char.TimeAxis" name="Char.TimeAxis" utype="spec:Char.TimeAxis">
<DESCRIPTION>Time Axis Characterization</DESCRIPTION>
<PARAM ID="TimeAxisName" datatype="char" name="TimeAxisName" utype="spec:Spectrum.Char.TimeAxis.Name" value="Time" arraysize="*">
<DESCRIPTION>Name for time axis</DESCRIPTION>
</PARAM>
<PARAM ID="TimeAxisUcd" datatype="char" name="TimeAxisUcd" utype="spec:Spectrum.Char.TimeAxis.Ucd" value="time" arraysize="*">
<DESCRIPTION>UCD for time</DESCRIPTION>
</PARAM>
<PARAM ID="TimeAxisUnit" datatype="char" name="TimeAxisUnit" utype="spec:Spectrum.Char.TimeAxis.Unit" value="s" arraysize="*">
<DESCRIPTION>Unit for time</DESCRIPTION>
</PARAM>
<PARAM ID="TimeLocation" unit="d" datatype="double" name="TimeLocation" ucd="time.epoch" utype="spec:Spectrum.Char.TimeAxis.Coverage.Location.Value"
  value="51999.54882939815"><DESCRIPTION>Midpoint of exposure on MJD scale</DESCRIPTION>
</PARAM>
<PARAM ID="TimeExtent" datatype="double" name="TimeExtent" ucd="time.duration;obs.exposure" utype="spec:Spectrum.Char.TimeAxis.Coverage.Bounds.Extent" value="2701">
<DESCRIPTION>Total exposure time</DESCRIPTION>
</PARAM>
<PARAM ID="TimeStart" datatype="double" name="TimeStart" ucd="time.start;obs.exposure" utype="spec:Spectrum.Char.TimeAxis.Coverage.Bounds.Start"
 value="51999.52502210648"><DESCRIPTION>Start time</DESCRIPTION>
</PARAM>
<PARAM ID="TimeStop" datatype="double" name="TimeStop" ucd="time.end;obs.exposure" utype="spec:Spectrum.Char.TimeAxis.Coverage.Bounds.Stop"
 value="51999.56039247685"><DESCRIPTION>Stop time</DESCRIPTION>
</PARAM>
</GROUP>
<GROUP ID="Char.FluxAxis" name="Char.FluxAxis" utype="spec:Char.FluxAxis">
<DESCRIPTION>Flux Axis Characterization</DESCRIPTION>
<PARAM ID="FluxAxisName" datatype="char" name="FluxAxisName" utype="spec:Spectrum.Char.FluxAxis.Name" value="Flux" arraysize="*">
<DESCRIPTION>Name for flux</DESCRIPTION>
</PARAM>
<PARAM ID="FluxAxisUcd" datatype="char" name="FluxAxisUcd" utype="spec:Spectrum.Char.FluxAxis.Ucd" value="phot.flux.density;em.wl" arraysize="*">
<DESCRIPTION>UCD for flux</DESCRIPTION>
</PARAM>
<PARAM ID="FluxAxisUnit" datatype="char" name="FluxAxisUnit" utype="spec:Spectrum.Char.FluxAxis.Unit" value="10**(-17) erg/cm**2/s/angstrom" arraysize="*">
<DESCRIPTION>Unit for flux</DESCRIPTION>
</PARAM>
<PARAM ID="FluxStatError" datatype="double" name="FluxStatError" ucd="stat.error;phot.flux" utype="spec:Spectrum.Char.FluxAxis.Accuracy.StatError" value="0">
<DESCRIPTION>Flux statistical error</DESCRIPTION>
</PARAM>
<PARAM ID="FluxCalibration" datatype="char" name="FluxCalibration" utype="spec:Spectrum.Char.FluxAxis.Accuracy.Calibration" value="Absolute" arraysize="*">
<DESCRIPTION>Type of flux calibration</DESCRIPTION>
</PARAM>
</GROUP>
```

```
<GROUP ID="Data.SpectralAxis" name="Data.SpectralAxis" utype="spec:Data.SpectralAxis">
<DESCRIPTION>Spectral Axis Data</DESCRIPTION>
<FIELDref ref="DataSpectralValue"/>
<FIELDref ref="DataSpectralBinLow"/>
<FIELDref ref="DataSpectralBinHigh"/>
<PARAM ID="DataSpectralUcd" datatype="char" name="DataSpectralUcd" utype="spec:Spectrum.Data.SpectralAxis.Ucd" value="em.wl" arraysize="*">
<DESCRIPTION>UCD for spectral coord</DESCRIPTION>
</PARAM>
<PARAM ID="DataSpectralUnit" datatype="char" name="DataSpectralUnit" utype="spec:Spectrum.Data.SpectralAxis.Unit" value="angstrom" arraysize="*">
<DESCRIPTION>Unit for spectral coord</DESCRIPTION>
</PARAM>
</GROUP>
<GROUP ID="Data.FluxAxis" name="Data.FluxAxis" utype="spec:Data.FluxAxis">
<DESCRIPTION>Flux Axis Data</DESCRIPTION>
<FIELDref ref="DataFluxValue"/>
<FIELDref ref="DataFluxStatErrLow"/>
<FIELDref ref="DataFluxStatErrHigh"/>
<FIELDref ref="DataFluxQuality"/>
<PARAM ID="DataFluxUcd" datatype="char" name="DataFluxUcd" utype="spec:Spectrum.Data.FluxAxis.Ucd" value="phot.flux.density;em.wl" arraysize="*">
<DESCRIPTION>UCD for flux</DESCRIPTION>
</PARAM>
<PARAM ID="DataFluxUnit" datatype="char" name="DataFluxUnit" utype="spec:Spectrum.Data.FluxAxis.Unit" value="10**(-17) erg/cm**2/s/angstrom" arraysize="*">
<DESCRIPTION>Unit for flux</DESCRIPTION>
</PARAM>
</GROUP>


<DATA>
<TABLEDATA>
<TR><TD>3822.0819974268243</TD><TD>3821.6419893046409</TD><TD>3822.5220562097306</TD><TD>0.6790500283241272</TD><TD>0</TD><TD>0</TD><TD>83886080</TD></TR>
<TR><TD>3822.9621656591976</TD><TD>3822.5220562097306</TD><TD>3823.40232578105</TD><TD>0.67889797687530518</TD><TD>0</TD><TD>0</TD><TD>83886080</TD></TR>
<TR><TD>3823.8425365811263</TD><TD>3823.40232578105</TD><TD>3824.2827980652619</TD><TD>0.67874801158905029</TD><TD>0</TD><TD>0</TD><TD>83886080</TD></TR>
<TR><TD>3824.7231102392952</TD><TD>3824.2827980652619</TD><TD>3825.1634731090558</TD><TD>0.67860102653503418</TD><TD>0</TD><TD>0</TD><TD>83886080</TD></TR>
<TR><TD>3825.6038866803838</TD><TD>3825.1634731090558</TD><TD>3826.0443509591169</TD><TD>0.67845600843429565</TD><TD>0</TD><TD>0</TD><TD>83886080</TD></TR>
<TR><TD>3826.4848659510972</TD><TD>3826.0443509591169</TD><TD>3826.9254316621559</TD><TD>0.67831301689147949</TD><TD>0</TD><TD>0</TD><TD>83886080</TD></TR>
<TR><TD>3827.3660480981362</TD><TD>3826.9254316621559</TD><TD>3827.8067152648787</TD><TD>0.6781730055809021</TD><TD>0</TD><TD>0</TD><TD>16777216</TD></TR>
<TR><TD>3828.2474331682283</TD><TD>3827.8067152648787</TD><TD>3828.6882018140186</TD><TD>0.67803502082824707</TD><TD>0</TD><TD>0</TD><TD>16777216</TD></TR>
...
<TR><TD>9200.25786648233</TD><TD>9199.1987086227764</TD><TD>9201.3171462889586</TD><TD>2.4964599609375</TD><TD>0</TD><TD>0</TD><TD>16777216</TD></TR>
<TR><TD>9202.37654805671</TD><TD>9201.3171462889586</TD><TD>9203.4360717996115</TD><TD>2.4953100681304932</TD><TD>0</TD><TD>0</TD><TD>16777216</TD></TR>
<TR><TD>9204.4957175317122</TD><TD>9203.4360717996115</TD><TD>9205.55548526707</TD><TD>2.4940199851989746</TD><TD>0</TD><TD>0</TD><TD>16777216</TD></TR>
<TR><TD>9206.61537501971</TD><TD>9205.55548526707</TD><TD>9207.6753868036922</TD><TD>2.4925899505615234</TD><TD>0</TD><TD>0</TD><TD>16777216</TD></TR>
<TR><TD>9208.7355206330667</TD><TD>9207.6753868036922</TD><TD>9209.7957765218853</TD><TD>2.4910099506378174</TD><TD>0</TD><TD>0</TD><TD>16777216</TD></TR>
</TABLEDATA>
</DATA>
</TABLE>
</RESOURCE>
</VOTABLE>
```

# Part 4: FITS serialization

# 9 FITS serialization

## 9.1 Mapping Spectrum to FITS

**FITS serialization design:** We define a reference serialization of this data model as a FITS binary table. The table represents a spectrum or photometry point as a single row of a table.

This serialization is a special case of an SED (or spectral association) serialization which uses one row per spectral segment; in that case, variable-length arrays may be used to contain the array quantities. In each case below where a 'variable length array' is specified, fixed length arrays are suitable for a single spectrum or for multiple spectra where all the arrays are the same length, but readers should be prepared to handle the variable length case.

For SEDs, another approach would be to have one FITS HDU per spectrum or photometry point. However this was rejected as unworkable, as the overhead of 5760 bytes (2 FITS blocks) per photometry point would inflate the data for the photometry-only SED case by factors of around 50-100.

**Standard table keywords:** In table F.1 we give the mapping of data model fields to FITS columns and keywords. For each column, the standard keywords TTYPEn, TUNITn, TFORMn should be provided. Order of keywords and columns is not significant, except that it is strongly recommended that RA and Dec be in adjacent columns or keywords. Additional keywords and columns which are not part of the model (including other conventions such as e.g. TDMINn) are allowed to be present, but are not guaranteed to be propagated by VO software.

**Keywords and columns: 'Greenbank' convention** In Table F.1 we give single metadata items as keywords; arrays of data (members of the Spectrum.Data classes) are stored as columns, and Table F.1 gives the column name, i.e. the value of the keyword TTYPEn. The 'Source' column in Table F.1 indicates if the name (if keyword) or value (if column) is a FITS standard (S), an existing convention (C) such as one of the HEA conventions, or is newly invented (N).

In some cases, the column data arrays may have the same value for each data point. In this case we may use the 'Greenbank' convention in which the column is omitted and replaced by a keyword whose name is the same as the column. Further, in SED applications when multiple spectrum data lines are present, some metadata may differ from line to line and be promoted from keyword to column. Therefore, implementors should check both keywords and column names for the appropriate tokens.

**TUTYP and TUCD keywords:** We map the FITS columns to the model by using TUTYPn keywords. TUTYPn (string-valued) gives the data model field name (UTYPE string) for the data in column n. Thus, the x and y axes (i.e. spectral coordinate and flux-like axes) of the spectrum have TUTYPn value of Spectrum.Data.SpectralAxis.Value and Spectrum.Data.FluxAxis.Value respectively.

Different kinds of x and y axis are identified by the Spectrum.Data.SpectralAxis.UCD and Spectrum.Data.FluxAxis.UCD data model fields, which are mapped to TUCDn keywords. TUCDn (string valued) gives the UCD corresponding to the data in column n. Both TUTYPn and TUCDn should be present for any column which corresponds to a Spectrum data model field; they are optional for any additional data columns which are not part of the Spectrum model. The units of spectral coordinate and flux are given in the TUNITn keys of the corresponding data columns. There is no separate provision for units of Char.SpectralAxis or Char.FluxAxis; these are required to be the same as for the data.

The TTYPEn keywords for the x and y columns are free, but it is strongly recommended that (for consistency of style with WCS Paper 3) the values for the x axis have for their first 4 characters 'WAVE', 'FREQ' and 'ENER' for the case of wavelength, frequency and energy respectively. We also recommend the value 'FLUX' for the y axis, where appropriate. Nevertheless, it is the TUTYPn and TUCDn keywords that should be used to interpret the semantics of the file.

**WAVE, ENER, and FREQ** In the header metadata, such as the Spectrum.Char entries, we use SPEC_ keywords to denote the spectral axis generically, but in the table columns (Spectrum.Data entries) we use the terms WAVE_, ENER_, and FREQ_ as appropriate. Thus if the Spectrum.Data.SpectralAxis.Value field is WAVE, the SpectralAxis.Accuracy.BinLow field should be WAVE_LO; if Value is FREQ, BinLow should be FREQ_LO. We believe the small extra parsing overhead is worth it for the readability and interoperability (since these names have been used in existing FITS files) of the crucial main data table.

**Char and Data keywords:** The model contains both Characterization metadata, giving overall typical values for quantities such as spectral resolution, and the Data object, which can include such quantities on a per-pixel basis. In some cases, the FITS serialization allows the same token for both Char (as a keyword) and Data (as a column name). The name, unit and UCD fields for Char.FluxAxis and Char.SpectralAxis are required to be the same as for Data.FluxAxis and Data.SpectralAxis. The case of TimeAxis is a little different, since there may be no Data.TimeAxis present, and there exist already some HEA conventions for recording TimeAxis characterization, notably the TIMEUNIT keyword. Note that TIMESYS, if present, must be TT.

**VOCLASS keyword:** We add a new keyword VOCLASS to describe the VO object represented by the FITS table. The value of VOCLASS should be 'SPECTRUM 1.00'.

**WCS table keywords:** The spectral coordinate may also be identified by optional 1Sn_1 and 1CTYPn keywords as per WCS Paper 3. Table 9 of that paper implies that each data column which is a function of the spectral coord needs a pair of such keywords. Applications which implement the spectrum data model may ignore the WCS keys and interpret the file by recognizing 'by spec' (using TUTYPn) which column is the spectral coordinate and that FLUX, etc. are functions of it, but the WCS keys give a general FITS application a chance at making sense of the file. In the example, TTYPE5='ERR_LO' and TUTYP5='Spectrum.Data.FluxAxis.Accuracy.StatErrLow'; the WCS keyword 1CTYPE5='WAVE-TAB' indicates that the data in column 5 is a function of wavelength, and that the wavelengths are in a lookup table. The WCS keyword 1S5_1='WAVE' indicates that the lookup table for the x-axis of column 5 (in this case, the wavelengths that the ERR_LO values correspond to) is in the column with TTYPEn='WAVE', in this case column 1.

Note that APERTURE has also been used elsewere as string-valued to indicate a named aperture; this is not allowed here.

The mid-exposure value is a required field for the internal data model; however it can be calculated from TSTART and TSTOP if they are present, and is then optional for the FITS serialization. The dataset start and stop wavelength may be provided in standard FITS as TDMINn/TDMAXn where n is the number of the column with the wavelengths.

For FITS, CoordSys.SpaceFrame.ucd is required to be the same as Char.SpatialAxis.ucd and CoordSys.TimeFrame.ucd is required to be equal to the default value "time". CoordSys.SpectralFrame.ucd is required to be the same as Data.SpectralAxis.ucd, present as a TUCDn value.

To express the CoordSys.RedshiftFrame, we recommend using a FITS WCS system with suffix 'Z' applied to the spectral coordinate axis, when appropriate. CoordSys.RedshiftFrame.DopplerDefinition is represented by the first four characters of TCTYPnZ and should have the values VRAD, VOPT, ZOPT or VELO, as per the convention for spectral CTYPE keywords in Paper III of the FITS WCS system. CoordSys.RedshiftFrame.RefPos is represented by SPECSYSZ and should have values as listed in Paper III of the FITS WCS system.

Table F.1: FITS keywords for VO Spectrum

| Data model field | FITS keyword | Source | Value if fixed |
|---|---|---|---|
| DataModel | VOCLASS | N | SPECTRUM 1.0 |
| Length | DATALEN | N | |
| Type | VOSEGT | N | |
| CoordSys.ID | VOCSID | N | |
| CoordSys.SpaceFrame.Name | RADECSYS | S | e.g. ICRS or FK5 |
| CoordSys.SpaceFrame.Equinox | EQUINOX | S | e.g. 2000.0 |
| CoordSys.SpaceFrame.ucd | SKY_UCD | | pos.eq |
| CoordSys.SpaceFrame.RefPos | SKY_REF | S | |
| CoordSys.TimeFrame.Name | TIMESYS | C | TT |
| CoordSys.TimeFrame.ucd | - | C | time |
| CoordSys.TimeFrame.Zero | MJDREF | C | default 0.0 |
| CoordSys.TimeFrame.RefPos | | | (not used) |
| CoordSys.SpectralFrame.RefPos | SPECSYS | S | (see below) |
| CoordSys.SpectralFrame.ucd | TUCDn | C | = Data.SpectralAxis.ucd |
| CoordSys.SpectralFrame.Redshift | REST_Z | N | |
| CoordSys.SpectralFrame.Name | SPECNAME | | |
| CoordSys.RedshiftFrame.Name | ZNAME | | |
| CoordSys.RedshiftFrame.DopplerDefinition | TCTYPnZ | | |
| CoordSys.RedshiftFrame.RefPos | SPECSYSZ | S | |
| Curation.Publisher | VOPUB | N | |
| Curation.Reference | VOREF | N | |
| Curation.PublisherID | VOPUBID | N | |
| Curation.Version | VOVER | N | |
| Curation.ContactName | CONTACT | N | |
| Curation.ContactEmail | EMAIL | N | |
| Curation.Rights | VORIGHTS | N | |
| Curation.Date | VODATE | N | |
| Curation.PublisherDID | DS_IDPUB | N | |
| Target.Name | OBJECT | S | |
| Target.Description | OBJDESC | N | |
| Target.Class | SRCCLASS | N | |
| Target.spectralClass | SPECTYPE | N | |
| Target.redshift | REDSHIFT | C | |
| Target.pos | RA_TARG, DEC_TARG | C | |
| Target.VarAmpl | TARGVAR | N | |
| DataID.Title | TITLE | C | |
| DataID.Creator | AUTHOR | S | |
| DataID.Collection | COLLECTn | N | |
| DataID.DatasetID | DS_IDENT | N | |
| DataID.CreatorDID | CR_IDENT | N | |
| DataID.Date | DATE | S | |
| DataID.Version | VERSION | C | |
| DataID.Instrument | INSTRUME | S | |
| DataID.CreationType | CRETYPE | N | |
| DataID.Logo | VOLOGO | N | |
| DataID.Contributor | CONTRIBn | N | |
| DataID.DataSource | DSSOURCE | N | |
| DataID.Bandpass | SPECBAND | N | |

| Data model field | FITS keyword | Source | Value if fixed |
|---|---|---|---|
| Derived.SNR | DER_SNR | N | |
| Derived.redshift.value | DER_Z | N | |
| Derived.redshift.statError | DER_ZERR | N | |
| Derived.redshift.Confidence | DER_ZCNF | N | |
| Derived.VarAmpl | DER_VAR | N | |
| TimeSI | TIMESDIM | N | |
| SpectralSI | SPECSDIM | N | |
| FluxSI | FLUXSDIM | N | |
| Omitted Char fields, values inherited from Spectrum.Data | | | |
| Char.FluxAxis.Name | - | - | TTYPEn for FLUX |
| Char.FluxAxis.Unit | - | - | Same as Data |
| Char.FluxAxis.ucd | - | - | Same as Data |
| Char.SpectralAxis.Name | - | - | Same as Data |
| Char.SpectralAxis.Unit | - | - | Same as Data |
| Char.SpectralAxis.ucd | - | - | Same as Data |
| Char.TimeAxis.Name | - | - | TIME |
| Char.TimeAxis.ucd | - | - | time |
| Char.SpatialAxis.Name | - | | (not used) |
| Char.SpatialAxis.Unit | - | | deg |
| Char Fields which are the same as for Spectrum.Data | | | |
| Char.FluxAxis.Accuracy.StatError | STAT_ERR | C | |
| Char.FluxAxis.Accuracy.SysError | SYS_ERR | C | |
| Char.TimeAxis.Accuracy.StatError | TIME_ERR | N | |
| Char.TimeAxis.Accuracy.SysError | TIME_SYE | N | |
| Char.TimeAxis.Resolution | TIME_RES | N | |

| Data model field | FITS keyword | Source | Value if fixed |
|---|---|---|---|
| | Char Fields which are only present in Char | | |
| Char.FluxAxis.Calibration | FLUX_CAL | N | |
| Char.SpectralAxis.Calibration | SPEC_CAL | N | |
| Char.SpectralAxis.Coverage.Location.Value | SPEC_VAL | N | |
| Char.SpectralAxis.Coverage.Bounds.Extent | SPEC_BW | N | |
| Char.SpectralAxis.Coverage.Bounds.Start | TDMINn | | |
| Char.SpectralAxis.Coverage.Bounds.Stop | TDMAXn | | |
| Char.SpectralAxis.SamplingPrecision. SamplingPrecisionRefVal.FillFactor | SPEC_FIL | N | |
| Char.SpectralAxis.SamplingPrecision. SampleExtent | SPEC_BIN | N | |
| Char.SpectralAxis.Accuracy.BinSize | SPEC_BIN | N | |
| Char.SpectralAxis.Accuracy.StatError | SPEC_ERR | N | |
| Char.SpectralAxis.Accuracy.SysError | SPEC_SYE | N | |
| Char.SpectralAxis.Resolution | SPEC_RES | N | |
| Char.SpectralAxis.ResPower | SPEC_RP | N | |
| Char.SpectralAxis.Coverage.Support.Extent | SPECWID | N | |
| Char.TimeAxis.Unit | TIMEUNIT | C | |
| Char.TimeAxis.Accuracy.BinSize | TIMEDEL | C | |
| Char.TimeAxis.Calibration | TIME_CAL | N | |
| Char.TimeAxis.Coverage.Location.Value | TMID | N | |
| Char.TimeAxis.Coverage.Bounds.Extent | TELAPSE | C | |
| Char.TimeAxis.Coverage.Bounds.Start | TSTART | C | |
| Char.TimeAxis.Coverage.Bounds.Stop | TSTOP | C | |
| Char.TimeAxis.Coverage.Support.Extent | EXPOSURE | C | |
| Char.TimeAxis.SamplingPrecision. SamplingPrecisionRefVal.FillFactor | DTCOR | C | |
| Char.TimeAxis.SamplingPrecision. SampleExtent | TIMEDEL | S | |
| Char.SpatialAxis.ucd | SKY_UCD | N | pos.eq |
| Char.SpatialAxis.Accuracy.StatErr | SKY_ERR | N | |
| Char.SpatialAxis.Accuracy.SysError | SKY_SYE | N | |
| Char.SpatialAxis.Calibration | SKY_CAL | N | |
| Char.SpatialAxis.Resolution | SKY_RES | N | |
| Char.SpatialAxis.Coverage.Location.Value | RA, DEC, etc. | C | |
| Char.SpatialAxis.Coverage.Bounds.Extent | APERTURE | C | |
| Char.SpatialAxis.Coverage.Support.Area | REGION | N | String value |
| Char.SpatialAxis.Coverage.Support.Extent | AREA | N | |
| Char.SpatialAxis.SamplingPrecision. SamplingPrecisionRefVal.FillFactor | SKY_FILL | N | |
| Char.SpatialAxis.SamplingPrecision. SampleExtent | TCDLTn | S | |

| Data model field | FITS keyword | Source | Value if fixed |
|---|---|---|---|
| | | | |
| | Per-data-point values | | |
| Data.FluxAxis.Value | TTYPEn | S | FLUX |
| UTYPE of above ... | TUTYPn | N | 'Spectrum.Data.FluxAxis.Value' |
| Data.FluxAxis.Unit | TUNITn | S | |
| Data.FluxAxis.ucd | TUCDn | N | (same as Char) |
| Data.FluxAxis.Accuracy.StatError | TTYPEn | N | ERR |
| Data.FluxAxis.Accuracy.StatErrLow | TTYPEn | C | ERR_LO |
| Data.FluxAxis.Accuracy.StatErrHigh | TTYPEn | C | ERR_HI |
| Data.FluxAxis.Accuracy.SysError | TTYPEn | C | SYS_ERR |
| Data.FluxAxis.Quality | TTYPEn | C | QUALITY |
| Data.FluxAxis.QualityN | TTYPEn | C | QUALn |
| Data.SpectralAxis.Value | TTYPEn | S | WAVE,ENER,FREQ |
| UTYPE of above ... | TUTYPn | N | 'Spectrum.Data.SpectralAxis.Value' |
| Data.SpectralAxis.Unit | TUNITn | S | (same as Char) |
| Data.SpectralAxis.ucd | TUCDn | N | (same as Char) |
| Data.SpectralAxis.Accuracy.BinSize | TTYPEn | N | WAVE_BIN,ENER_BIN, FREQ_BIN |
| Data.SpectralAxis.Accuracy.BinLow | TTYPEn | N | WAVE_LO,ENER_LO, FREQ_LO |
| Data.SpectralAxis.Accuracy.BinHigh | TTYPEn | N | WAVE_HI,ENER_HI, FREQ_HI |
| Data.SpectralAxis.Accuracy.StatError | TTYPEn | N | WAVE_ERR,ENER_ERR, FREQ_ERR |
| Data.SpectralAxis.Accuracy.StatErrLow | TTYPEn | N | WAVE_ELO,ENER_ELO, FREQ_ELO |
| Data.SpectralAxis.Accuracy.StatErrHigh | TTYPEn | N | WAVE_EHI,ENER_EHI, FREQ_EHI |
| Data.SpectralAxis.Accuracy.SysError | TTYPEn | N | WAVE_SYE,ENER_SYE, FREQ_SYE |
| Data.SpectralAxis.Resolution | TTYPEn | N | WAVE_RES,ENER_RES, FREQ_RES |
| Data.TimeAxis.Value | TTYPEn | C | TIME |
| UTYPE of above ... | TUTYPn | N | 'Spectrum.Data.TimeAxis.Value' |
| Data.TimeAxis.Unit | TUNITn | S | (same as Char) |
| Data.TimeAxis.ucd | TUCDn | N | time |
| Data.TimeAxis.Accuracy.BinLow | TTYPEn | N | TIME_LO |
| Data.TimeAxis.Accuracy.BinHigh | TTYPEn | N | TIME_HI |
| Data.TimeAxis.Accuracy.BinSize | TIMEDEL | S | |
| Data.TimeAxis.Resolution | TTYPEn | N | TIME_RES |
| Data.TimeAxis.Accuracy.StatError | TTYPEn | N | TIME_ERR |
| Data.TimeAxis.Accuracy.StatErrLow | TTYPEn | N | TIME_ELO |
| Data.TimeAxis.Accuracy.StatErrHigh | TTYPEn | N | TIME_EHI |
| Data.TimeAxis.Accuracy.SysError | TTYPEn | N | TIME_SYE |
| Data.BackgroundModel.Value | TTYPEn | N | BGFLUX |
| UTYPE of above ... | TUTYPn | N | 'Spectrum.Data.BackgroundModel.Value' |
| Data.BackgroundModel.Unit | TUNITn | S | (same as FluxAxis) |
| Data.BackgroundModel.ucd | TUCDn | N | (same as FluxAxis) |
| Data.BackgroundModel.Accuracy.StatError | TTYPEn | N | BG_ERR |
| Data.BackgroundModel.Accuracy.StatErrLow | TTYPEn | N | BG_ELO |
| Data.BackgroundModel.Accuracy.StatErrHigh | TTYPEn | N | BG_EHI |
| Data.BackgroundModel.Accuracy.SysError | TTYPEn | N | BG_SYE |
| Data.BackgroundModel.Quality | TTYPEn | N | BGQUAL |

## 9.2 Expressing the spectrum spatial coordinates in FITS

FITS has a sophisticated mechanism for expressing celestial coordinates. However, it applies only to image axes or table columns. If you want to express a single celestial position in the header of a FITS binary table, the WCS conventions do not apply. You could add an extra pair of columns to the table giving the same position in each row, but that would be wasteful.

Here we propose a local convention leveraging the existing WCS conventions:

- The keyword names for the coordinates are those used in the first four characters of the CTYPE values for the WCS paper: e.g. RA, DEC, GLON, GLAT.
- The coordinate system is identified by the keyword SKY_UCD with values such as pos.eq, etc.
- The RADECSYS and EQUINOX keywords should be used when appropriate.
- Values are always in degrees.
- The VOCSID optional keyword is provided to allow VO coordinate system names used in Spectrum and STC to be propagated to FITS. Its value is not relevant in the FITS context.

## 9.3 The SPECSYS keyword

We note the allowed values of the SPECSYS keyword from Greisen et al and the corresponding values from the VO STC:

| FITS | STC | Meaning |
|---|---|---|
| TOPOCENT | TOPOCENTER | Topocenter |
| GEOCENTR | GEOCENTER | Geocenter |
| BARYCENT | BARYCENTER | Solar System Barycenter |
| HELIOCEN | HELIOCENTER | Heliocenter |
| LSRK | LSRK | Kinematic local standard of rest |
| LSRD | LSRD | Dynamic local standard of rest |
| GALACTOC | GALACTIC_CENTER | Galactic center |
| LOCALGRP | LOCAL_GROUP_CENTER | Local group barycenter |
| CMBDIPOL | - | Frame of the Cosmic Microwave Background dipole |
| SOURCE | - | Source rest frame |

## 9.4 An instance example

We summarize this with a sample FITS extension header.

```
XTENSION= 'BINTABLE'           / binary table extension
BITPIX  =                    8 / 8-bit bytes
NAXIS   =                    2 / 2-dimensional binary table
NAXIS1  =                57344 / width of table in bytes
NAXIS2  =                    1 / number of rows in table
PCOUNT  =                    0 / size of special data area
GCOUNT  =                    1 / one data group (required keyword)
TFIELDS =                    7 / number of fields in each row
EXTNAME = 'SPECTRUM '           / name of this binary table extension
VOCLASS = 'Spectrum V1.0'      / VO Data Model
DATALEN =       180            / Segment size
VOSEGT  = 'Spectrum'           / Segment type
VOCSID  = 'MY-ICRS-TOPO'          / Coord sys ID
RADECSYS= 'FK5    '           / Not default - usually ICRS
EQUINOX =  2.0000000000000E+03 / default
TIMESYS = 'TT     '           / Time system
MJDREF  =  0.0                 / [d] MJD zero point for times
SPECSYS = 'TOPOCENT'            / Wavelengths are as observed
VOPUB   = 'CfA Archive'        / VO Publisher authority
VOREF   = '2006ApJ...999...99X'  / Bibcode for citation
VOPUBID = 'ivo://cfa.harvard.edu' / VO Publisher ID URI
VOVER   = '1.0'                   / VO Curation version
CONTACT = 'Jonathan McDowell, CfA'/
EMAIL   = 'jcm@cfa.harvard.edu'   /
VORIGHTS= 'public'  /
VODATE  = '2004-08-30'     /
DS_IDPUB= 'ivo://cfa.harvard.edu/spec#10304' / Publisher DID for dataset
COMMENT  DS_IDPUB usually the same as DS_IDENT?
OBJECT  = 'ARP 220 '           / Source name
OBJDESC = 'Merging galaxy Arp 220' / Source desc
SRCCLASS= 'Galaxy'             /
SPECTYPE= 'ULIRG'              /
REDSHIFT=             0.01812  / Emission redshift
RA_TARG      =     233.73791700     / [deg] Observer's specified target RA
DEC_TARG     =      23.50333300     / [deg] Observer's specified target Dec
TARGVAR =    0.2    /  20 percent variability amplitude
TITLE   = 'Observations of Merging Galaxies' /
AUTHOR  = 'MMT Archive'       / VO Creator
COLLECT1= 'Misc Pointed Observations'  / Collection
DS_IDENT= 'ivo://cfa.harvard.edu/spec#10304' / Publisher DID for dataset
CR_IDENT= 'ivo://cfa.harvard.edu/tdc#MMT4302-102' / Creator internal ID for dataset
DATE    = '2004-08-30T14:18:17' / Date and time of file creation
VERSION =  2                     / Reprocessed 2004 Aug
TELESCOP= 'MMT '              / Telescope  [Not part of Spectrum DM]
INSTRUME= 'MMT/BCS '          / Instrument
FILTER  = 'G220    '          / Grating  [Not part of Spectrum DM]
CRETYPE = 'Archival'          / Not an on-the-fly dataset
VOLOGO  = 'http://cfa.harvard.edu/vo/cfalogo.jpg' / VO Creator logo
CONTRIB1= 'Jonathan McDowell'  / Contributor
CONTRIB2= 'Wilhelm Herschel'  / Contributor
CONTRIB3= 'Harlow Shapley'     / Contributor
DSSOURCE= 'Pointed'   /  Survey or pointed, etc
DER_SNR =         5.0  /   Estimate of signal-to-noise
DER_Z   =       0.01845   /  Redshift measured in this spectrum
DER_ZERR =       0.00010   /  Error in DER_Z
```

```
TIMESDIM= 'T'                  / Time SIDim
SPECSDIM= '10-10 L'             / Spectral SIDim
FLUXSDIM= '10+7 ML-1T-3'       / Flux SDim
SYS_ERR =        0.05            / Fractional systematic error in flux
FLUX_CAL= 'Calibrated'   /
SPEC_ERR=        0.01 / Stat error in spec coord, in SPEC units
SPEC_SYE=        0.001   /   Frac sys error in spec coord
SPEC_CAL= 'Calibrated'
SPEC_RES=              5.0  / [angstrom] Spectral resolution
SPECBAND= 'Optical'    / SED.Bandpass
SPEC_RP =            800.0  / Spectral resolving power
SPEC_VAL=           4100.0  / [angstrom]  Characteristic spec coord
SPEC_BW =           1800.0  / [angstrom]  Width of spectrum
SPEC_FIL=              1.0  / No gaps between channels
TIME_CAL = 'Calibrated'   /
DATE-OBS= '2004-06-03T21:18:17' / Date and time of observation
EXPOSURE =   1500.015         / [s] Effective exposure time
TSTART   =  52984.301203      / [d] MJD
TSTOP    =  52984.318564      / [d] MJD
TMID     =  52984.309883     / [d] MJD mid expsoure
SKY_CAL  = 'Calibrated'  /
SKY_RES =              1.0  / [arcsec] Spatial.Resolution
RA      =     233.73791       / [deg] Pointing position
DEC     =      23.50333       / [deg] Pointing position
APERTURE=       2.0           / [arcsec]  Aperture diameter/Slit width
TIME     =  52984.309883      / [d] MJD of midpoint


COMMENT  ---------------------------
COMMENT  WCS Paper 3 Keywords
1S4_1    = 'WAVE'            / Column name with spectral coord
1CTYP4   = 'WAVE-TAB'        / Spectral coord is WAVE
1S5_1    = 'WAVE'            / Column name with spectral coord
1CTYP5   = 'WAVE-TAB'        / Spectral coord is WAVE
1S6_1    = 'WAVE'            / Column name with spectral coord
1CTYP6   = 'WAVE-TAB'        / Spectral coord is WAVE
1S7_1    = 'WAVE'            / Column name with spectral coord
1CTYP7   = 'WAVE-TAB'        / Spectral coord is WAVE
COMMENT  ---------------------------
TTYPE1 = 'WAVE'   / Wavelength
TFORM1 = '180E'
TUNIT1 = 'angstrom'
TUCD1  = 'em.wl'                     /
TDMIN1 = 3195.0 /
TDMAX1 = 5005.0 /
TUTYP1 = 'Spectrum.Data.SpectralAxis.Value'
TTYPE2 = 'WAVE_LO' /
TFORM2 = '180E'
TUNIT2 = 'angstrom'
TUTYP2 = 'Spectrum.Data.SpectralAxis.Accuracy.StatErrLow'
TTYPE3 = 'WAVE_HI' /
TFORM3 = '180E'
TUNIT3 = 'angstrom'
TUTYP3 = 'Spectrum.Data.SpectralAxis.Accuracy.StatErrHigh'
TTYPE4 = 'FLUX' /
TFORM4 = '180E'
TUNIT4 = 'erg cm**(-2) s**(-1) angstrom**(-1)'
TUTYP4 = 'Spectrum.Data.FluxAxis.Value'
TUCD4  = 'phot.fluDens;em.wl'     / Type of Y axis: F-lambda
```

```
TTYPE5 = 'ERR_LO' /
TFORM5 = '180E'
TUNIT5 = 'erg cm**(-2) s**(-1) angstrom**(-1)'
TUTYP5 = 'Spectrum.Data.FluxAxis.Accuracy.StatErrLow'
TTYPE6 = 'ERR_HI' /
TFORM6 = '180E'
TUNIT6 = 'erg cm**(-2) s**(-1) angstrom**(-1)'
TUTYP6 = 'Spectrum.Data.FluxAxis.Accuracy.StarErrHigh'
TTYPE7 = 'QUALITY' /
TFORM7 = '180I'
TUTYP7 = 'Spectrum.Data.FluxAxis.Quality'
```

The data would look like

```
WAVE   WAVE_LO WAVE_HI FLUX  ERR_LO  ERR_HI  QUALITY
3200.0 3195.0 3205.0 1.48E-12 2.0E-14 2.0E-14   0
3210.0 3205.0 3215.0 1.52E-12 3.0E-14 3.0E-14   0
3220.0 3215.0 3225.0 0.38E-12 0.38E-12 0.0      0
3230.0 3225.0 3235.0 1.62E-12 3.0E-14 3.0E-14   0
...
5000.0 4995.0 5005.0 1.33E-11 3.0E-13 3.0E-13   1
```